\documentclass[aps,prl,twocolumn,10pt]{revtex4-2}

\usepackage{mathrsfs}
\usepackage{bbm}
\usepackage{amsmath}
\usepackage{amssymb}
\usepackage{amsthm}
\usepackage{graphicx}
\usepackage{MnSymbol}
\usepackage{mathtools}
\usepackage[italicdiff]{physics}
\usepackage{dsfont}
\usepackage{bbold}

\makeatletter
\usepackage{hyperref}
\usepackage{color}
\definecolor{supcol}{RGB}{10,50,180}
\definecolor{eqcol}{RGB}{220,10,100}
\hypersetup{
	colorlinks,
	citecolor=supcol,
	linkcolor=eqcol,
	urlcolor=supcol
}

\allowdisplaybreaks

\DeclareMathOperator{\Var}{Var}

\newcommand{\mca}{\mathcal}
\newcommand{\mbb}{\mathbb}
\newcommand{\mrm}{\mathrm}

\newcommand{\mds}{\mathds}

\newcommand{\sectionprl}[1]{{\em #1}\/---}

\begin{document}
\title{Thermodynamic Uncertainty Relation with Quantum Feedback}

\author{Ryotaro Honma}
\email{ryotaro.honma@yukawa.kyoto-u.ac.jp}
\affiliation{Center for Gravitational Physics and Quantum Information, Yukawa Institute for Theoretical Physics, Kyoto University, Kitashirakawa Oiwakecho, Sakyo-ku, Kyoto 606-8502, Japan}

\author{Tan Van Vu}
\email{tan.vu@yukawa.kyoto-u.ac.jp}
\affiliation{Center for Gravitational Physics and Quantum Information, Yukawa Institute for Theoretical Physics, Kyoto University, Kitashirakawa Oiwakecho, Sakyo-ku, Kyoto 606-8502, Japan}

\date{\today}

\begin{abstract}
Fluctuations are intrinsic to microscopic systems and impose fundamental limits on nonequilibrium precision, as captured by the thermodynamic uncertainty relation (TUR), which links current fluctuations to entropy production. While feedback control is expected to further suppress fluctuations, its role within the TUR framework has remained unclear, particularly in quantum systems where control is inherently information-driven. In this Letter, we consider open quantum systems weakly coupled to a thermal environment, in which quantum jumps are continuously monitored, and Markovian feedback is applied. Using quantum mutual information to quantify the information contribution induced by feedback, we derive a finite-time TUR for arbitrary time-integrated currents in terms of entropy production and mutual information. Our results uncover how feedback control suppresses fluctuations together with thermodynamic cost and establish a fundamental precision bound imposed by information-based control. As an application, we analyze a quantum clock model and demonstrate that the clock precision can be enhanced by feedback control in the presence of a single thermal reservoir.
\end{abstract}

\maketitle

\sectionprl{Introduction}How are fluctuations in microscopic systems constrained by thermodynamics? Recent advances in stochastic thermodynamics have uncovered fundamental limits governing such fluctuations. Since fluctuations are inherently prominent at the microscopic scale, their suppression constitutes a central problem in nonequilibrium thermodynamics. A key result in this context is the thermodynamic uncertainty relation (TUR) \cite{Barato.2015.PRL,Gingrich.2016.PRL,Horowitz.2020.NP}, which establishes a trade-off between the relative fluctuation of a current---quantified by the variance divided by the squared mean---and the total entropy production. The TUR thus provides a fundamental principle that fluctuations cannot be reduced without incurring thermodynamic cost and has far-reaching implications for nonequilibrium physics \cite{Pietzonka.2018.PRL,Hartich.2021.PRL,Manikandan.2020.PRL,Vu.2020.PRE}.

Beyond passive thermodynamic constraints, feedback control \cite{Wiseman.2009} is expected to further suppress fluctuations \cite{Potts.2019.PRE,Vu.2020.JPA}. In quantum systems, feedback control has demonstrated significant advantages across a broad range of applications \cite{Zhang.2017.PR}, including work extraction \cite{Elouard.2017.PRL,Masuyama.2018.NC}, enhanced metrological precision \cite{Fallani.2022.PRXQ,Salvia.2023.PRL}, state stabilization \cite{Vijay.2012.N,Campagne-Ibarcq.2013.PRX,Rossi.2018.N}, squeezing \cite{Wiseman.1994.PRA2}, entanglement manipulation \cite{Carvalho.2007.PRA}, and error correction \cite{Ahn.2002.PRA,Livingston.2022.NC}. These processes can be naturally framed within information thermodynamics \cite{Parrondo.2015.NP}, in which desired control tasks are achieved at the expense of information-related costs. Accordingly, several formulations of the second law of thermodynamics incorporating information costs have been established for measurement and feedback processes in quantum systems. For discrete measurement and feedback protocols, such formulations include those based on the quantum-classical (QC) mutual information \cite{Sagawa.2008.PRL}, coarse-grained entropy production \cite{Prech.2024.PRL}, and mutual information for Markovian feedback control \cite{Strasberg.2017.PRX} and cyclic heat engines \cite{Koshihara.2022.PRE}. For continuous-time non-Markovian feedback described by stochastic master equations, a generalized second law employing QC transfer entropy has also been developed \cite{Yada.2022.PRL}.

\begin{figure}[b]
	\centering
	\includegraphics[width=0.9\linewidth]{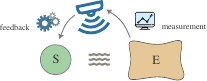}
	\protect\caption{Schematic illustration of an open quantum system $S$ under Markovian feedback control. Quantum jumps induced by interactions with the environment $E$ are continuously monitored, and feedback control is applied instantaneously based on the detected quantum-jump outcomes.}\label{fig:setup}
\end{figure}

In contrast to these advances in second-law-type inequalities, the relationship between current fluctuations and information-thermodynamic costs in the presence of feedback control remains poorly understood \cite{QTRoadmap.2026.QST}. Despite numerous extensions of the TUR to quantum systems \cite{Guarnieri.2019.PRR,Carollo.2019.PRL,Hasegawa.2020.PRL,Miller.2021.PRL.TUR,Vu.2022.PRL.TUR,Prech.2025.PRL,Vu.2025.PRXQ,Macieszczak.2024.arxiv,Kwon.2025.CP,Moreira.2025.PRE,Palmqvist.2025.PRL,Blasi.2026.PRL,Brandner.2025.PRL,Vu.2026.PRL,Honma.2026.PRA}, only a few studies have addressed the role of feedback control. Among these few works, a kinetic uncertainty relation for continuous measurement and Markovian unitary feedback has been derived \cite{Yunoki.2025.arxiv}; however, the resulting bound involves dynamical activity rather than entropy production or information. Another approach derives a generalized TUR from fluctuation theorems applicable to systems under generic feedback control \cite{Potts.2019.PRE}. While this result applies to broad classes of protocols, its generality comes at the expense of tightness: the relative fluctuation of the current is bounded only \emph{exponentially} by the thermodynamic cost. Moreover, the inequality involves quantities defined in the time-reversed (backward) process, which obscures its physical interpretation.

In this Letter, we bridge this gap for open quantum systems undergoing Markovian feedback control. We consider a general system weakly coupled to a thermal environment, in which quantum jumps are continuously monitored, and feedback is applied instantaneously (see Fig.~\ref{fig:setup}). Using quantum mutual information, we quantify the information exploited by feedback control that is relevant to current fluctuations and derive a second law of thermodynamics for measurement and feedback processes. As our main result, we establish a finite-time TUR for arbitrary time-integrated currents in terms of entropy production and mutual information [Eq.~\eqref{eq:main.result}]. This relation recovers the original TUR in the absence of feedback control and in the classical limit. Our findings quantitatively reveal how feedback control suppresses current fluctuations together with thermodynamic cost, thereby imposing a fundamental limitation set by feedback. We apply our framework to a quantum clock model, demonstrating how feedback control enhances clock operation even with a single thermal reservoir.

\sectionprl{Setup}We consider a continuous measurement and feedback process for a finite-dimensional system $S$ weakly coupled to a thermal environment $E$.
The system is continuously monitored through measurements of the environment, during which quantum jumps occur and are recorded in a memory $M$. 
Conditioned on the detected quantum-jump outcomes, feedback control is applied instantaneously to the system. 
We assume that, in the absence of feedback control, the system dynamics is governed by the Gorini-Kossakowski-Sudarshan-Lindblad (GKSL) master equation \cite{Lindblad.1976.CMP,Gorini.1976.JMP}.
Such measurement-and-feedback schemes have been experimentally realized on various platforms, including superconducting circuits \cite{Campagne-Ibarcq.2013.PRX,Minev.2019.N,Chen.2021.PRL} and Rydberg atoms \cite{Sayrin.2011.N}.

To describe the underlying stochastic dynamics, it is convenient to consider a short-time interval $dt$ and take the continuous-time limit $ dt\to 0$ at the end \cite{fnt1}.
At each time $t$, the system either undergoes an abrupt quantum jump, determined by one of the jump operators $\{L_k\}_{k\ge 1}$, or evolves smoothly under nonunitary dynamics generated by the effective Hamiltonian $H_{\rm eff}\coloneqq H-(i/2)\sum_{k\ge 1}L_k^\dagger L_k$, where $H$ denotes the system Hamiltonian.
Whenever the $k$th jump occurs, a completely positive trace-preserving map $\mca{F}_k$ is applied instantaneously to the system as feedback control. 
The conditioned system state then becomes $\mca{F}_k[M_k\varrho_tM_k^\dagger]/\tr\mca{F}_k[M_k\varrho_tM_k^\dagger]$, where $M_k=L_k\sqrt{ dt}$ for $k\ge 1$.
When no jump occurs, no feedback is applied and the system state evolves to $M_0\varrho_tM_0^\dagger/\tr(M_0\varrho_tM_0^\dagger)$, with $M_0=\mbb{1}-iH_{\rm eff} dt$, where $\mbb{1}$ denotes the identity operator.
At the ensemble level, the unconditional state of the system is updated after each time step as
\begin{equation}
	\varrho_{t+ dt}=M_0\varrho_tM_0^\dagger+\sum_{k\ge 1}\mca{F}_k[M_k\varrho_tM_k^\dagger].
\end{equation}
Taking the $ dt\to 0$ limit yields the following master equation \cite{Wiseman.1994.PRA}:
\begin{align}
	\dot{\varrho}_t &=-i[H,\varrho_t]+\sum_{k\geq 1}\qty(\mca{F}_k[L_k\varrho_t L_k^\dagger] -\frac{1}{2}\{L_k^\dagger L_k,\varrho_t \})\notag\\
	&\eqqcolon \mca{L}^{(\mrm{fb})}[\varrho_t].
\end{align}
Hereafter, we set $\hbar=k_B=1$ for simplicity.
For clarity of exposition, we focus on time-independent Hamiltonians and jump operators, although our analysis extends straightforwardly to general time-dependent driving.
To ensure thermodynamic consistency, we assume that the jump operators satisfy the local detailed balance condition \cite{Horowitz.2013.NJP,Manzano.2018.PRX,Vu.2023.PRX}. 
Specifically, the jump operators come in pairs $(k,k^*)$ such that $L_k=e^{\Delta s_k/2}L_{k^*}^\dagger$, where $\Delta s_k$ denotes the environmental entropy change associated with the $k$th jump.
This structure allows for an unambiguous definition of entropy production, admitting a meaningful decomposition into system and environmental contributions, as well as a detailed fluctuation theorem for stochastic entropy production \cite{Horowitz.2013.NJP}. 
Finally, we exclusively consider unital feedback maps $\{\mca{F}_k\}_{k\ge 1}$ (i.e., $\mca{F}_k[\mbb{1}]=\mbb{1}$).
This class includes relevant operations such as unitary transformations, projective measurements, and identity maps.

The underlying stochastic dynamics over a fixed duration $\tau$ gives rise to an ensemble of stochastic trajectories, each denoted by $\Gamma=\{(t_1,k_1),\dots,(t_N,k_N)\}$, where the $k_i$th jump occurs at a random time $t_i$ for $i=1,\dots,N$.
The total number of quantum jumps $N$ associated with each trajectory is itself a stochastic variable.
For an individual trajectory $\Gamma$, a time-integrated current is defined as
\begin{equation}
	J(\Gamma)\coloneqq\sum_{i=1}^Nc_{k_i}.
\end{equation}
Here, $\{c_k\}_{k\ge 1}$ are counting coefficients that satisfy the antisymmetry condition $c_k=-c_{k^*}$.
In other words, whenever the $k$th jump is detected, the weight $c_k$ is added to the current $J$.
Relevant examples include heat flux in heat engines, particle currents in transport setups, and tick currents in clocks.
In general, all moments of the current $J$ can be computed using the method of full counting statistics \cite{Landi.2024.PRXQ}.
Defining the generating function $Z(u)\coloneqq \operatorname{tr}[e^{\mca{L}_u^{\rm (fb)}\tau}(\varrho_0)]$, we obtain the $n$th moment of $J$ as $\ev{J^n}=\eval{(-i\partial_u)^n Z (u)}_{u=0}$, where the tilted superoperator $\mca{L}_u^{\rm (fb)}$ is given by
\begin{align}
	\mca{L}_u^{(\mrm{fb})}(\circ)&\coloneqq 
    \mca{L}^{(\mrm{fb})}(\circ)+\sum_{k\geq 1}(e^{iuc_{k}}-1)\mca{F}_k[L_{k}\circ L_{k}^\dagger].
\end{align}
We are particularly interested in the relationship between the relative fluctuation $\Var[J]/\ev{J}^2$ and information-thermodynamic costs, where $\Var[J]\coloneqq\ev{J^2}-\ev{J}^2$ denotes the current variance.

We now discuss the information-thermodynamic framework of the measurement and feedback process described above.
Hereafter, the superscript $-$ denotes pre-feedback quantities, whereas post-feedback quantities are written without a superscript.
The conditioned state immediately after the $k$th jump is given by $\varrho_{t+ dt,k}^-= \mca{M}_k[\varrho_t]/p_k$, where $\mca{M}_k[\circ]= M_k\circ M_k^\dagger$ and $p_k= \tr(\mca{M}_k[\varrho_t])$.
Accordingly, the composite state of the system and the memory before feedback can be written as $\varrho_{t+ dt,\rm tot}^- = \sum_{k\ge 0}\mca{M}_k[\varrho_t]\otimes\dyad{k}$, where $\{\ket{k}\}_{k\ge 0}$ denotes a computational basis of the memory.
Similarly, the conditioned state immediately after feedback for the $k$th jump is $\varrho_{t+ dt,k}= \mca{F}_k[\varrho_{t+ dt,k}^-]$, and the corresponding ensemble state reads $\varrho_{t+ dt,\rm tot}= \sum_{k\ge 0}\mca{F}_k[\mca{M}_k(\varrho_t)]\otimes\dyad{k}$.
Here, we introduce the identity map $\mca{F}_0[\circ]\coloneqq\circ$ for notational convenience.
With these definitions, we can show that the total entropy production before feedback is always nonnegative \cite{Supp.PhysRev},
\begin{align}
	S(\varrho_{t+ dt}^-)-S(\varrho_t)+ dS^{\rm env}\geq 0.\label{eq:ent.inc.meas}
\end{align}
Here, $\varrho_{t+ dt}^-\coloneqq \tr_M\varrho_{t+ dt,\mrm{tot}}^-$ is the unconditional system state immediately before feedback, $S(\varrho)\coloneqq -\tr(\varrho\ln\varrho)$ denotes the von Neumann entropy, and $dS^{\rm env}$ is the average entropy production of the environment over the infinitesimal time interval $dt$, given by 
\begin{align}
	dS^{\rm env}\coloneqq  dt \sum_{k\geq 1} \Delta s_k \tr(L_k \varrho_t L_k^\dagger).\label{eq:ent.env.inc.meas}
\end{align}
Since $\tr(L_k\varrho_tL_k^\dagger)$ gives the rate at which the jump $L_k$ occurs, Eq.~\eqref{eq:ent.env.inc.meas} can be understood as the average of the entropy increment $\Delta s_k$, weighted by the probability of each jump event \cite{fnt2}.
Furthermore, since entropy does not decrease under unital maps, the inequality
\begin{align}
	S(\varrho_{t+ dt,k})-S(\varrho_{t+ dt,k}^{-})\geq 0 \label{eq:ent.inc.unit.map}
\end{align}
holds for each $k$th jump event.
Indeed, a unital feedback map is both necessary and sufficient for Eq.~\eqref{eq:ent.inc.unit.map} to hold for all input states, and equality is achieved for unitary feedback (i.e., $\mca{F}_k[\circ]=U_k \circ U_k^\dagger$ with a unitary operator $U_k$).
It is worth noting that, although the entropy does not decrease during feedback for each individual stochastic jump, the entropy of the system may nevertheless decrease at the ensemble level.

It is important to quantify the amount of information exploited during the measurement and feedback process.
To this end, the quantum mutual information between the system $S$ and the memory $M$ is introduced, defined as $I(S:M)\coloneqq S(\varrho_S)+S(\varrho_M)-S(\varrho_{\mrm{tot}})$, where $\varrho_{S(M)}\coloneqq \tr_{M(S)}\varrho_{\mrm{tot}}$.
Using this definition, the mutual information between the system and the memory immediately after measurement and immediately after feedback, corresponding to the composite states $\varrho_{t+ dt,\rm tot}^-$ and $\varrho_{t+ dt,\rm tot}$, respectively, is given by
\begin{align}
	I_{t+ dt}^-=S(\varrho_{t+ dt}^-)-\sum_{k\geq 0}p_k S(\varrho_{t+ dt,k}^-),\label{eq:mi.meas}\\
	I_{t+ dt}=S(\varrho_{t+ dt})-\sum_{k\geq 0}p_k S(\varrho_{t+ dt,k}).\label{eq:mi.fb}
\end{align}
The difference between these two quantities, which quantifies the amount of information exploited during the feedback step, is defined as
\begin{equation}\label{eq:mi.def}
	dI\coloneqq I_{t+ dt} - I_{t+ dt}^-.
\end{equation}
Using the monotonicity of entropy under unital channels in Eq.~\eqref{eq:ent.inc.unit.map}, we find that this informational quantity provides a lower bound on the entropy change \cite{Supp.PhysRev},
\begin{equation}
	S(\varrho_{t+dt})-S(\varrho_{t+dt}^-)\ge dI.\label{eq:ent.inc.fb}
\end{equation}
This inequality implies that the system entropy may decrease, but that its reduction cannot exceed the amount of mutual information consumed during feedback. This is also consistent with the standard picture of Maxwell's demon in information thermodynamics \cite{Parrondo.2015.NP}. Another proof that exploits the invariance of the memory entropy under feedback is given in Appendix A.

\sectionprl{Main results}With the key quantities defined above, we are now ready to present our results.
First, we derive a second law of thermodynamics for measurement and feedback control, identifying the information-thermodynamic cost that is relevant to the precision of currents. 
By combining Eqs.~\eqref{eq:ent.inc.meas} and \eqref{eq:ent.inc.fb}, we obtain
\begin{equation}
	dS^{\rm sys}+ dS^{\rm env}- dI\geq 0,\label{eq:main.SL}
\end{equation}
where $dS^{\rm sys}\coloneqq S(\varrho_{t+ dt})-S(\varrho_t)$ denotes the entropy change of the system.
With the total entropy production defined as $dS^{\rm tot}\coloneqq dS^{\rm sys}+dS^{\rm env}$, the inequality \eqref{eq:main.SL} implies that the entropy change $dS^{\rm tot}$ is constrained by the mutual information $dI$ exploited during the feedback step. 
In other words, the total entropy may decrease after measurement and feedback control, but only by an amount bounded by the mutual information.

It is worth noting that the formulation \eqref{eq:main.SL} of the second law of thermodynamics is tighter than the existing result \cite{Strasberg.2017.PRX}, which applies to more general settings than the present framework.
Specifically, the second law for the measurement step derived in Ref.~\cite{Strasberg.2017.PRX} reads
\begin{equation}\label{eq:ent.inc.meas.prx}
	S(\varrho_{t+dt}^-)-S(\varrho_t)+H(p)-I_{t+dt}^-+dS^{\rm env}\ge 0,
\end{equation}
where $H(p)\coloneqq -\sum_{k\ge 0}p_k\ln p_k$ is the Shannon entropy.
The second law associated with the feedback step coincides with Eq.~\eqref{eq:ent.inc.fb} when no heat is dissipated during feedback.
By invoking the inequality on the entropy of a mixture of quantum states \cite{Nielsen.2000},
\begin{equation}
	S(\varrho_{t+dt}^-)\le \sum_{k\ge 0}p_kS(\varrho_{t+dt,k}^-)+H(p),
\end{equation}
one obtains $I_{t+dt}^-\le H(p)$, which readily shows that the inequality \eqref{eq:ent.inc.meas} is tighter than Eq.~\eqref{eq:ent.inc.meas.prx}.
Consequently, the second law \eqref{eq:main.SL} provides a more refined constraint on the irreversibility of the full measurement and feedback process than the extant result.

In the continuous-time limit, the second law \eqref{eq:main.SL} takes the differential form $\dot\Sigma\coloneqq\dot S^{\rm tot}-\dot I\ge 0$, where $\dot X\coloneqq\lim_{dt\to 0} dX/dt$, and its time-integrated version reads $\Sigma=S^{\rm tot}-I\ge 0$ (see Appendix B for the explicit form of $\dot I$).
Using a quantum-information-theoretic approach, we prove that the relative fluctuation of any current is constrained by both entropy production and the information exploited during feedback as
\begin{equation}\label{eq:main.result}
	\frac{\Var[J]}{\ev{J}^2} \geq \frac{2(1 + \delta_{J})^2}{\Sigma}.
\end{equation}
This constitutes our main result and quantitatively reveals the crucial role of information in reducing current fluctuations, even in the presence of negative entropy production.
It holds for arbitrary durations and initial states; the proof is deferred to Appendix C.
Here, $\delta_J$ is a correction term, explicitly defined as $\delta_J\coloneqq\ev{J}_\varphi/\ev{J}$, where $\ev{J}_\varphi\coloneqq\int_0^\tau\dd{t}\sum_{k\geq 1}c_{k}\tr(L_{k}\varphi_t L_{k}^\dagger)$ and the traceless operator $\varphi_t$ evolves according to
\begin{gather}
	\dot{\varphi}_t=\mca{L}^{\mrm{(fb)}}[\varphi_t]+\sum_{k\geq 1}\ell_k(t)\qty(\mca{F}_k[L_k\varrho_tL_k^\dagger]-\frac{1}{2}\{L_k^\dagger L_k,\varrho_t\}),\notag\\
	\ell_k(t)\coloneqq \frac{\tr(L_k \varrho_t L_k^\dagger) - \tr(L_{k^*} \varrho_t L_{k^*}^\dagger)}{\tr(L_k \varrho_t L_k^\dagger) + \tr(L_{k^*} \varrho_t L_{k^*}^\dagger)},\label{eq:phi.dyn}
\end{gather}
with the initial condition $\varphi_0=\mbb{0}$.

Several remarks on the result \eqref{eq:main.result} are in order.
First, although evaluating $\delta_J$ generally requires detailed knowledge of the dynamics, this does not limit the bound's operational use. Analogously to the original TUR \cite{Barato.2015.PRL}, it can be used to infer the otherwise inaccessible dissipation from the directly measured current and its relative fluctuation. In the short-time limit, $\delta_J=O(\tau)\to 0$ \cite{Supp.PhysRev}, so the correction drops out and the inequality reduces to an experimentally accessible, tomography-free bound on the steady-state rate $\dot\Sigma$,
\begin{equation}
    \dot\Sigma\ge\lim_{\tau\to 0}\frac{2\ev{J}^2}{\tau\Var[J]}.
\end{equation}
Beyond this practical role, $\delta_J$ quantifies the deviation from the original TUR induced by quantum effects and feedback. Even without feedback, it can remain finite in the steady state \cite{Vu.2025.PRXQ}, a genuinely quantum contribution irreducible to simpler quantities such as the coherence of $\varrho$ \cite{Kalaee.2021.PRE}. It also provides insights into how strongly current fluctuations can be suppressed at a given cost: $|1+\delta_J|\ll 1$ signals that high precision is attainable at small $\Sigma$, beyond the reach of the original TUR.
Second, by incorporating dynamical activity into the lower bound, a tighter TUR can be obtained \cite{Vo.2022.JPA,Vu.2025.PRXQ}. 
Specifically, defining the average number of quantum jumps during the time interval $\tau$ as $A\coloneqq\int_0^\tau\dd{t}\sum_{k\ge 1}\tr(L_k\varrho_tL_k^\dagger)$, we can strengthen the bound \eqref{eq:main.result} to
\begin{equation}\label{eq:tkur}
	\frac{\Var[J]}{\ev{J}^2} \geq (1 + \delta_{J})^2\frac{4A}{\Sigma^2}\Phi\qty(\frac{\Sigma}{2A})^2,
\end{equation}
where $\Phi$ denotes the inverse function of $x\tanh(x)$ and satisfies $\Phi(x)\ge\max(\sqrt{x},x)$.
Last, our result provides a quantitative framework for analyzing the performance of thermal machines in the presence of feedback control from the perspective of precision-cost trade-offs. 
For instance, as a corollary, we can show that the precision of a clock is fundamentally limited by the information-thermodynamic cost per tick and quantum coherence \cite{Honma.2026.PRA}.

\begin{figure}[t]
	\centering
	\includegraphics[width=1\linewidth]{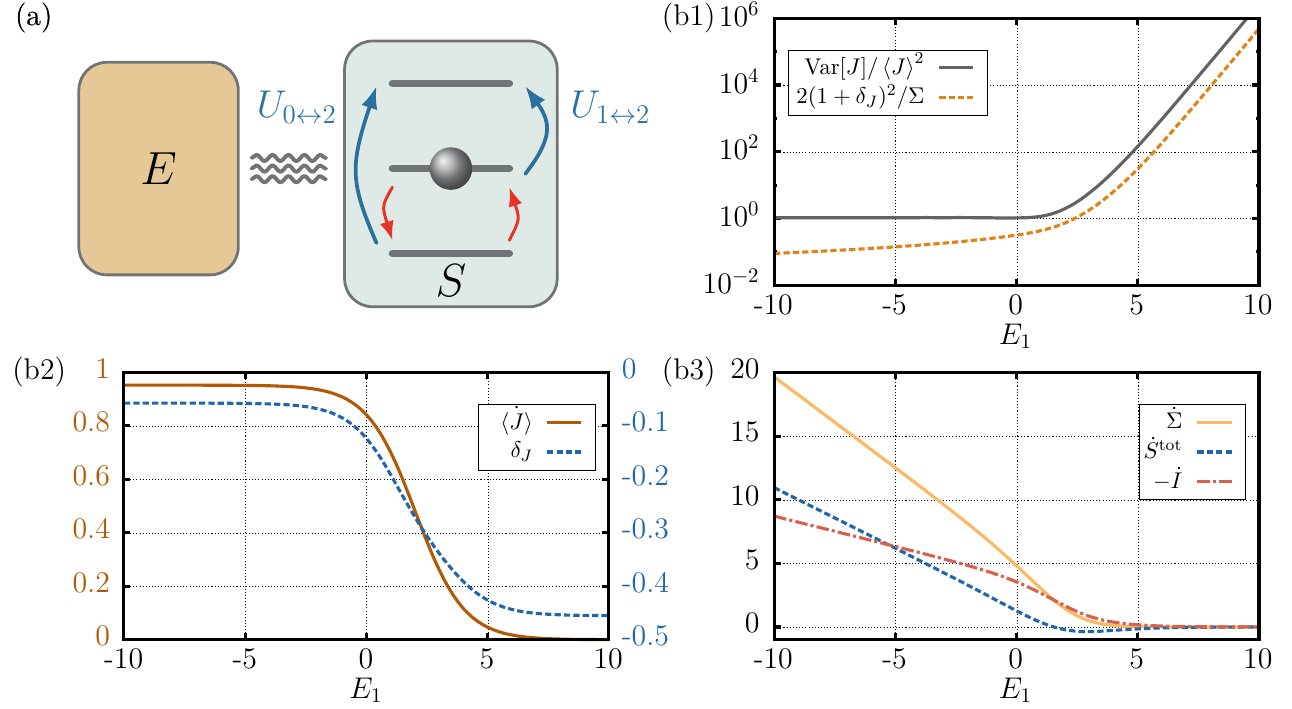}
	\protect\caption{(a) Schematic illustration of the quantum clock, composed of a three-level system weakly coupled to a thermal reservoir. (b) Numerical verification of the quantum TUR \eqref{eq:main.result} with feedback in the stationary state. The vertical axis in (b1) is shown on a logarithmic scale; all other axes are linear.
(b1) Solid and dashed lines represent the relative fluctuation $\Var[J]/\ev{J}^2$ and its lower bound $2(1+\delta_{J})^2/\Sigma$, respectively.
(b2) Solid and dashed lines depict the average current rate $\dot{\ev{J}}$ and the correction term $\delta_{J}$, respectively.
(b3) Solid, dashed, and dash-dotted lines represent $\dot\Sigma$, $\dot S^{\rm tot}$, and $-\dot I$, respectively.
The energy level $E_1$ of the state $\ket{1}$ is varied over the range $[-10,10]$, while the remaining parameters are fixed as $E_0=0$, $E_2=1$, $\beta =1$, $\gamma_{1\rightarrow 0}=10$, $\gamma_{2\rightarrow 1}=0.5$, $\gamma_{2\rightarrow 0}=1$, and $\tau = 1$.}\label{fig:clock}
\end{figure}

\sectionprl{Example}We demonstrate our results using a simple quantum clock model, illustrated in Fig.~\ref{fig:clock}(a).
The clock consists of a three-level system with states $\ket{0}$, $\ket{1}$, and $\ket{2}$.
The Hamiltonian is given by $H=\sum_{i=0}^{2}E_i \dyad{i}$, where $E_i$ denotes the energy level of the state $\ket{i}$.
In general, the energy levels are not required to satisfy the ordering $E_0<E_1<E_2$.
The system is weakly coupled to a thermal reservoir at temperature $T=1/\beta$, which mediates transitions between each pair of states.
The jump operator describing the transition from $\ket{i}$ to $\ket{j}$ is given by $L_{i\rightarrow j}=\sqrt{\gamma_{i\rightarrow j}}\dyad{j}{i}$, where $\gamma_{i\rightarrow j}$ denotes the corresponding transition rate.
These rates satisfy the detailed balance condition $\gamma_{i\rightarrow j}=e^{-\beta (E_j -E_i)}\gamma_{j\rightarrow i}$.
%In the absence of feedback, the system relaxes toward the thermal equilibrium state $e^{-\beta H}/\tr (e^{-\beta H})$, and no steady current can be generated.
In the absence of feedback, the system relaxes toward the thermal equilibrium state $e^{-\beta H}/\tr(e^{-\beta H})$. This follows from the local detailed balance condition together with the relation $[H,L_{i\to j}]=(E_j-E_i)L_{i\to j}$, under which the Gibbs state is stationary and all net transition currents vanish.

To enable the system to operate as a clock even in the presence of a single reservoir, we apply unitary feedback conditioned on jump events.
Specifically, the unitary operations $U_{1\leftrightarrow 2}$ and $U_{0\leftrightarrow 2}$ are applied immediately after the jumps $L_{0\rightarrow 1}$ and $L_{1\rightarrow 0}$, respectively, while no feedback is applied for other jumps.
Here, $U_{i\leftrightarrow j} = \dyad{i}{j}+\dyad{j}{i}+\dyad{3-i-j}$, which swaps the states $\ket{i}$ and $\ket{j}$ while leaving the remaining state $\ket{3-i-j}$ unchanged.
Consequently, whenever a jump between $\ket{0}$ and $\ket{1}$ occurs, the feedback brings the system to the state $\ket{2}$.
Under this feedback scheme, the dynamics of the clock is governed by
\begin{align}
  \dot{\varrho}_{t}=-i[H, \varrho_t]
  &+\mca{D}[U_{1\leftrightarrow 2}L_{0\rightarrow 1}]\varrho_t
  +\mca{D}[U_{0\leftrightarrow 2} L_{1\rightarrow 0}]\varrho_t \notag\\
  &+\mca{D}[L_{0\rightarrow 2}]\varrho_t
  +\mca{D}[L_{2\rightarrow 0}]\varrho_t
  \notag \\
  &+\mca{D}[L_{1\rightarrow 2}]\varrho_t
  +\mca{D}[L_{2\rightarrow 1}]\varrho_t.
\end{align}
Here, $\mca{D}[L]\circ\coloneqq L\circ L^\dagger-\{L^\dagger L,\circ\}/2$.
Due to the unitary feedback, the system is driven out of thermal equilibrium and reaches a nonequilibrium stationary state. 
As a result, persistent currents can emerge in the system \cite{fnt3}.

In the following, we focus on the tick current $J$, which counts $+1$ for each jump from $\ket{2}$ to $\ket{0}$, $-1$ for each jump from $\ket{0}$ to $\ket{2}$, and $0$ for all other transitions.
This current quantifies the net number of ticks generated by the clock mechanism. 
To verify the main result \eqref{eq:main.result}, we analyze the steady state and numerically evaluate all relevant quantities. 
Motivated by experimental considerations, we consider the clock dynamics over a finite time interval $\tau$, rather than taking the asymptotic long-time limit.
Throughout the numerical calculations, we fix $E_0=0$ and $E_2=1$, and all energies are expressed in units of $E_2$. We vary the energy level $E_1$ of the state $\ket{1}$ over the range $[-10,10]$, spanning both the $E_1<0$ and $E_1>1$ regimes, while holding all other parameters fixed.
The numerical results are shown in Fig.~\ref{fig:clock}(b).

As illustrated in Fig.~\ref{fig:clock}(b2), a persistent tick current emerges due to the feedback.
The average current increases monotonically as $E_1$ decreases and saturates when $|E_1|$ becomes large; the correction term $\delta_J$ shows a similar trend.
Figure \ref{fig:clock}(b1) shows that the relative fluctuation $\Var[J]/\ev{J}^2$ of the tick current is consistently bounded from below by $2(1+\delta_{J})^2/\Sigma$, thereby numerically validating the main result \eqref{eq:main.result}.
Notably, the clock fluctuation is enhanced for large $E_1$, while remaining suppressed for small $E_1$.
Figure \ref{fig:clock}(b3) demonstrates that $\dot{\Sigma}=\dot{S}^{\rm tot}-\dot{I}$ is nonnegative for all parameters considered, confirming the validity of the second law \eqref{eq:main.SL} in this model.
Moreover, there exists a parameter regime in which the total entropy production rate $\dot{S}^{\rm tot}$ becomes negative, while the relative fluctuation of the tick current remains finite. 
This highlights the crucial role of the mutual information $I$ in enhancing the precision of currents.

\sectionprl{Conclusion}In this Letter, we established thermodynamic bounds for open quantum systems undergoing continuous measurement and Markovian feedback control. 
We first derived the second law of thermodynamics \eqref{eq:main.SL} for continuous measurement and feedback protocols, expressed in terms of entropy production and the mutual information between the system and the memory. 
Building on this result, we obtained the quantum TUR \eqref{eq:main.result} for arbitrary time-integrated currents, showing that, in addition to entropy production, the accumulated mutual information between the system and the memory plays a crucial role in suppressing relative current fluctuations. 
We numerically demonstrated our results using a minimal quantum clock model. 
Our findings provide a quantitative framework for understanding the fundamental limits of current precision in open quantum systems under continuous measurement and feedback control.

\sectionprl{Note added}Independently of our work, the authors of Ref.~\cite{Tojo.2026.arxiv} obtained related, yet distinct, results on quantum TURs under feedback control by incorporating quantum-classical-transfer entropy.

\begin{acknowledgments}
\sectionprl{Acknowledgments}We thank Kaito Tojo, Takahiro Sagawa, and Ken Funo for sharing their preliminary manuscript. This work was supported by JSPS KAKENHI Grants No.~JP23K13032, No.~JP26K00019, and No.~JP26H02015.
\end{acknowledgments}

\sectionprl{Data availability}The data are not publicly available upon publication. The data are available from the authors upon reasonable request.

% \bibliography{submission-final/refs_final}

\begin{thebibliography}{64}
\bibitem [{fnt({\natexlab{a}})}]{fnt1}%
  \BibitemOpen
  \href@noop {} {}Here, continuous monitoring obtains the quantum-jump record by detecting the environmental output after its interaction with the system, rather than subjecting the system to projective measurements at intervals $dt$. At fixed dynamical parameters, jump probabilities are linear in $dt$ to leading order, so taking $dt\to0$ preserves finite jump rates without inducing quantum-Zeno freezing.\BibitemShut {Stop}%
\bibitem [{fnt({\natexlab{b}})}]{fnt2}%
  \BibitemOpen
  \href@noop {} {}We note that, for global master equations satisfying the detailed balance condition (i.e., when $[H,L_k]=\Delta E_k L_k$, with $\Delta E_k$ being the energy change associated with jump $L_k$ and $\Delta s_k=-\Delta E_k/T$), the environmental entropy production in Eq.~\eqref{eq:ent.env.inc.meas} reduces exactly to the conventional thermodynamic form $dS^{\rm env}=-dQ/T$, where $dQ=\mathrm{tr}[H(\varrho_{t+dt}^- - \varrho_t)]$ denotes the heat absorbed by the system.\BibitemShut {Stop}%
\bibitem [{fnt({\natexlab{c}})}]{fnt3}%
  \BibitemOpen
  \href@noop {} {}We note that two-level quantum clocks with multiple reservoirs can also be considered, provided the mutual information term does not diverge; see Ref.~\cite{Supp.PhysRev} for further demonstrations. We adopt the present three-level system because it provides a simple clock that operates with a single reservoir and swap feedback while keeping the mutual information finite.\BibitemShut {Stop}%
\end{thebibliography}


\onecolumngrid
\begin{center}
    \textbf{End Matter}
\end{center}
\twocolumngrid

\sectionprl{Appendix A: Proof of Eq.~\eqref{eq:ent.inc.fb}}We show that Eq.~\eqref{eq:ent.inc.fb} can be interpreted as a reformulation of the second law for the joint system (system + memory).
Starting from the definition of the mutual information, we can decompose its change across the feedback step into the entropy changes of the system, the memory, and the joint system as
\begin{equation}
    dI=\delta S^{\mrm{sys}}+\delta S^{\mrm{mem}}-\delta S^{\mrm{joint}},
\end{equation}
where each entropy difference is evaluated between the states right before and after the feedback operation, defined by
\begin{align}
    \delta S^{\rm sys}\coloneqq S(\varrho_{t+dt})-S(\varrho_{t+dt}^-),\\
    \delta S^{\rm mem}\coloneqq S(\varrho_{t+dt,M})-S(\varrho_{t+dt,M}^-),\\
    \delta S^{\rm joint}\coloneqq S(\varrho_{t+dt,\rm tot})-S(\varrho_{t+dt,\rm tot}^-).
\end{align}
Here, $\varrho_{t+dt,M}^-=\tr_S\varrho_{t+dt,\rm tot}^-$ and $\varrho_{t+dt,M}=\tr_S\varrho_{t+dt,\rm tot}$.
The feedback operation, acting on the joint system, is described by the channel
\begin{equation}
    \mca{F}_{\rm fb}[\circ]=\sum_{k\ge 0}\mca{F}_k[\mel{k}{\circ}{k}]\otimes\dyad{k}.
\end{equation}
One readily verifies that $\varrho_{t+dt,\rm tot}=\mca{F}_{\rm fb}[\varrho_{t+dt,\rm tot}^-]$.
Moreover, $\mca{F}_{\rm fb}$ is a unital map, $\mca{F}_{\rm fb}[\mds{1}_S\otimes\mds{1}_M]=\mds{1}_S\otimes\mds{1}_M$, because each $\mca{F}_k$ is a unital map.
Since a unital map never decreases the entropy, we have $\delta S^{\rm joint}=S(\varrho_{t+dt,\rm tot}) - S(\varrho_{t+dt,\rm tot}^-) \ge 0$, which yields a second law for the joint system,
\begin{equation}
    \delta S^{\rm joint}=\delta S^{\rm sys} + \delta S^{\rm mem} - dI \ge 0.
\end{equation}
In our setup, the memory state remains unchanged during feedback. Hence, $\delta S^{\mrm{mem}}=0$, and the joint-system second law reduces immediately to $\delta S^{\mrm{sys}}-dI\ge 0$, which is precisely Eq.~\eqref{eq:ent.inc.fb}.

\sectionprl{Appendix B: Explicit form of the mutual information rate}Here we derive an explicit expression for the mutual information rate $\dot I$.
To this end, we evaluate $I_{t+dt}^-$ and $I_{t+dt}$ up to order $O(dt)$.
First, note that $p_k=\tr(L_k \varrho_t L_k^\dagger)dt$ for $k\geq 1$, $p_0=1-\sum_{k\ge 1}p_k$, and 
\begin{align}
    \varrho_{t+dt,0}^-&=\qty(\varrho_t+(-i[H,\varrho_t]- \sum_{k\geq 1} \{L_k^\dagger L_k ,\varrho_t \}/2)dt)/p_0,\\
    \varrho_{t+dt,k}^-&=L_k \varrho_t L_k^\dagger dt /p_k \quad (k\geq 1).
\end{align}
Using these expressions, we obtain
\begin{align}\label{eq:ent.conditioned}
	S(\varrho_{t+dt}^-)&=S(\varrho_t)-\tr(\mca{L}^{(0)}[\varrho_t]\ln \varrho_t)dt,
\end{align}
where the superoperator $\mca{L}^{(0)}$ is defined as $\mca{L}^{(0)}[\circ]\coloneqq -i[H,\circ]+\sum_{k\ge 1}\mca{D}[L_k]\circ$.
Substituting this into Eq.~\eqref{eq:mi.meas}, we obtain
\begin{equation}
  I_{t+dt}^- =\sum_{k\geq 1}p_k D\qty(\frac{L_k \varrho_t L_k^\dagger}{\tr[L_k \varrho_t L_k^\dagger]} \| \varrho_t),
\end{equation}
where $D(\varrho\|\phi)\coloneqq \tr[\varrho (\ln\varrho-\ln\phi)]$ is the relative entropy between states $\varrho$ and $\phi$.
Similarly, the mutual information after feedback can be written as
\begin{equation}
  I_{t+dt}=\sum_{k\geq 1}p_k D\qty(\frac{\mca{F}_k[L_k \varrho_t L_k^\dagger]}{\tr[L_k \varrho_t L_k^\dagger]} \| \varrho_t).
\end{equation}
Note that both $I_{t+dt}^-$ and $I_{t+dt}$ are $O(dt)$ quantities.
In the limit $dt\to 0$, the mutual information rate can be expressed explicitly as
\begin{equation}
	\dot I =\sum_{k\ge 1}\tr(L_k\varrho_tL_k^\dagger)\qty{D\qty(\mca{F}_k[\varrho_{t,k}'] \| \varrho_t) - D\qty(\varrho_{t,k}' \| \varrho_t)},
\end{equation}
where $\varrho_{t,k}'\coloneqq L_k \varrho_t L_k^\dagger/\tr(L_k \varrho_t L_k^\dagger)$.
It is worth noting that $\dot I$ may diverge if the support of any post-measurement state is not contained in that of the pre-measurement state.

\sectionprl{Appendix C: Proof sketch of the main result}We provide an outline of the proof of Eq.~\eqref{eq:main.result}; the detailed derivation is presented in the Supplemental Material \cite{Supp.PhysRev}.

From the inequality \eqref{eq:ent.inc.fb}, $S(\varrho_{t+dt})-S(\varrho_{t+dt}^-)-dI\geq 0$, it follows that
\begin{equation}\label{eq:ent.prod.lb1}
	dS^{\rm sys}+dS^{\mrm{env}}-dI\geq S(\varrho_{t+dt}^-)-S(\varrho_{t})+dS^{\mrm{env}}.
\end{equation}
In the limit $dt\to 0$, Eq.~\eqref{eq:ent.prod.lb1} becomes
\begin{equation}
	\dot\Sigma\ge -\tr(\mca{L}^{(0)}[\varrho_t]\ln \varrho_t)+\sum_{k\geq 1}\tr(L_k\varrho_t L_k^\dagger)\Delta s_k \eqqcolon\sigma.
\end{equation}
Using the spectral decomposition $\varrho_t =\sum_n p_n(t)\dyad{n_t}$ and defining $w^{mn}_{k}(t)\coloneqq|\mel{m_t}{L_{k}}{n_t}|^2$, one finds $\sigma=(1/2)\sum_{k\ge 1,m,n}\sigma_k^{mn}$ and $\dot A=(1/2)\sum_{k\ge 1,m,n}a_k^{mn}$, where
\begin{align}
	\sigma^{mn}_k & \coloneqq  (w^{mn}_{k} p_n-w^{nm}_{k^*} p_m)\ln \frac{w^{mn}_{k} p_n}{w^{nm}_{k^*} p_m},
  \\
	a^{mn}_k &\coloneqq w^{mn}_{k} p_n+w^{nm}_{k^*} p_m.
\end{align}
Next, we introduce a perturbed dynamics in which the jump operators are modified by a parameter $\theta$ as $L_{k,\theta}(t) = \sqrt{1 + \ell_{k}(t)\theta}L_{k}$ while the Hamiltonian and feedback channels $\{\mca{F}_k\}_{k\ge 1}$ remain unchanged.
For this measurement and feedback process, we can derive a generalized quantum Cram{\'e}r-Rao inequality for any time-integrated current $J$ as
\begin{equation}\label{eq:qCRcound}
  \frac{\Var[J]}{(\partial_\theta \ev{J}_\theta|_{\theta=0})^2}\geq \frac{1}{\mca{I}_0},
\end{equation}
where $\mca{I}_0$ denotes the quantum Fisher information evaluated at $\theta=0$, and $\ev{J}_\theta$ is the average current under the perturbed dynamics.
The quantum Fisher information can be evaluated as 
\begin{align}
  \mca{I}_0
  &=\int_0^\tau \dd{t} \sum_{k\geq 1}\ell_{k}^2 (t)\tr(L_{k}\varrho_t L_{k}^\dagger )\notag\\
  &=\int_0^\tau \dd{t} \frac{1}{2}\sum_{k\geq 1}\dfrac{ \qty[\sum_{m,n}(w_{k}^{mn}p_n-w_{k^*}^{nm}p_m)]^2}{\sum_{m,n}(w_{k}^{mn}p_n+w_{k^*}^{nm}p_m)}\notag\\
  &\leq \int_0^\tau \dd{t} \sum_{k\geq 1}\sum_{m,n}\frac{(\sigma^{mn}_k)^2}{8a^{mn}_k} \Phi \qty(\frac{\sigma^{mn}_k}{2a^{mn}_k})^{-2}\notag\\
  &\leq \int_0^\tau \dd{t} \frac{\sigma^2}{4\dot{A}} \Phi \qty(\frac{\sigma}{2\dot{A}})^{-2},\label{eq:FI.ub1}
\end{align}
where Jensen's inequality is applied to the concave function $(x^2/y)\Phi(x/y)^{-2}$.
Additionally, since $x^2\Phi(x)^{-2}$ is monotonically increasing for $x\ge 0$ and $\sigma\leq \dot{\Sigma}$, we obtain
\begin{equation}\label{eq:FI.ub2}
	\frac{\sigma^2}{4\dot{A}}\Phi\qty(\frac{\sigma}{2\dot{A}})^{-2}\leq\frac{\dot{\Sigma}^2}{4\dot{A}}\Phi\qty(\frac{\dot{\Sigma}}{2\dot{A}})^{-2}.
\end{equation}
Combining Eqs.~\eqref{eq:FI.ub1} and \eqref{eq:FI.ub2} and applying Jensen's inequality once more yields the following upper bound on the quantum Fisher information:
\begin{equation}
  \mca{I}_0 \leq \int_0^\tau \dd{t} \frac{\dot{\Sigma}^2}{4\dot{A}} \Phi \qty(\frac{\dot{\Sigma}}{2\dot{A}})^{-2}\leq \frac{\Sigma^2}{4A} \Phi \qty(\frac{\Sigma}{2A})^{-2}.\label{eq:Fi.tkur}
\end{equation}
The derivative of the current average can be calculated as
\begin{equation}\label{eq:pdv.ave.cur}
  \partial_\theta \ev{J}_\theta |_{\theta=0}=\ev{J}+\ev{J}_\varphi=\ev{J}(1+\delta_{J}).
\end{equation}
From Eq.~\eqref{eq:pdv.ave.cur}, $\ev{J}_\varphi$ can be identified as a contribution to the linear response of the current to the auxiliary parameter $\theta$. It may not, however, have an operational meaning as the response to an independently controllable physical drive, since $\theta$ is an auxiliary, theoretical construct rather than an experimental control. Finally, combining Eqs.~\eqref{eq:qCRcound}, \eqref{eq:pdv.ave.cur}, and \eqref{eq:Fi.tkur} yields the improved TUR \eqref{eq:tkur}:
\begin{equation}
  \frac{\Var[J]}{\ev{J}^2}\ge (1+\delta_{J})^2\frac{4A}{\Sigma^2}\Phi\qty(\frac{\Sigma}{2A})^2.
\end{equation}
The main result \eqref{eq:main.result} follows immediately from the inequality $\Phi (x)\geq \sqrt{x}$.

\end{document}

\title{Supplemental Material for \\``Thermodynamic Uncertainty Relation with Quantum Feedback''}

\author{Ryotaro Honma}
\email{ryotaro.honma@yukawa.kyoto-u.ac.jp}
\affiliation{Center for Gravitational Physics and Quantum Information, Yukawa Institute for Theoretical Physics, Kyoto University, Kitashirakawa Oiwakecho, Sakyo-ku, Kyoto 606-8502, Japan}

\author{Tan Van Vu}
\email{tan.vu@yukawa.kyoto-u.ac.jp}
\affiliation{Center for Gravitational Physics and Quantum Information, Yukawa Institute for Theoretical Physics, Kyoto University, Kitashirakawa Oiwakecho, Sakyo-ku, Kyoto 606-8502, Japan}

\begin{abstract}
This Supplemental Material provides detailed analytical calculations, including the derivation of the second law of thermodynamics under feedback, the thermodynamic uncertainty relation, and results for some special cases.
Equation and figure numbers are prefixed with S [e.g., Eq.~(S1) or Fig.~S1]. 
The numbers without this prefix [e.g., Eq.~(1) or Fig.~1] refer to the items in the main text.
\end{abstract}

\pacs{}
\maketitle

\tableofcontents

\section{Full counting statistics of currents}
Here, we derive the formulation of the $n$th moment of currents using the method of full counting statistics.

To this end, we discretize the time duration $\tau=N\Delta t$ and take the limit $N\to\infty$ at the end.
The system dynamics can be unraveled into an ensemble of stochastic trajectories $\Gamma=(k_1,\dots,k_N)$, where either the $k_j$th jump occurs, or the system undergoes a smooth nonunitary evolution (i.e., $k_j=0$) during the time interval $[(j-1)\Delta t,j\Delta t]$.
The probability of observing the stochastic trajectory $\Gamma$ is given by
\begin{align}
  p(\Gamma)=\tr[\mca{F}_{k_N}[M_{k_N}\dots M_{k_2}\mca{F}_{k_1}[M_{k_1}\varrho_0 M_{k_1}^\dagger]M_{k_2}^\dagger\dots M_{k_N}^\dagger ]],
\end{align}
where $M_k=L_k\sqrt{\Delta t}~\forall k\ge 1$ and $M_0=\mbb{1}-[iH+(1/2)\sum_{k\ge 1}L_k^\dagger L_k]\Delta t$.

All moments of the current $J$ can be computed using the generating function $Z(u)$ defined as follows:
\begin{equation}
    Z(u)\coloneqq\sum_{\Gamma}e^{iuJ(\Gamma)}p(\Gamma).
\end{equation}
It is evident that
\begin{equation}
    \ev{J^n}=\sum_{\Gamma}J(\Gamma)^np(\Gamma)=(-i\partial_u)^nZ(u)|_{u=0}.
\end{equation}
The generating function $Z(u)$ can be calculated as follows:
\begin{align}
    Z(u)&=\sum_{\Gamma}e^{iu\sum_{j=1}^Nc_{k_j}}\tr[\mca{F}_{k_N}[M_{k_N}\dots M_{k_2}\mca{F}_{k_1}[M_{k_1}\varrho_0 M_{k_1}^\dagger]M_{k_2}^\dagger\dots M_{k_N}^\dagger ]]\notag\\
    &=\sum_{\Gamma}\tr[e^{iuc_{k_N}}\mca{F}_{k_N}[M_{k_N}\dots M_{k_2}e^{iuc_{k_1}}\mca{F}_{k_1}[M_{k_1}\varrho_0 M_{k_1}^\dagger]M_{k_2}^\dagger\dots M_{k_N}^\dagger ]].
\end{align}
Now, we define the following tilted operators:
\begin{align}
    \varrho_{0,u}&\coloneqq\varrho_0,\\
    \varrho_{n\Delta t,u}&\coloneqq \sum_{k_1,\dots,k_n} e^{iuc_{k_n}}\mca{F}_{k_n}[M_{k_n}\dots M_{k_2}e^{iuc_{k_1}}\mca{F}_{k_1}[M_{k_1}\varrho_0 M_{k_1}^\dagger]M_{k_2}^\dagger\dots M_{k_n}^\dagger ].
\end{align}
It follows that $Z(u)=\tr\varrho_{\tau,u}$.
The tilted operator $\varrho_{n\Delta t,u}$ satisfies the following relation:
\begin{equation}
    \varrho_{n\Delta t,u}=\sum_{k\ge 0}e^{iuc_k}\mca{F}_k[M_k\varrho_{(n-1)\Delta t,u}M_k^\dagger].
\end{equation}
Taking the limit $\Delta t\to 0$ of this relation, we obtain the following differential equation for the tilted operator:
\begin{equation}
    \dot\varrho_{t,u}=\mca{L}_u^{\rm (fb)}(\varrho_{t,u}),
\end{equation}
where 
\begin{equation}
    \mca{L}_u^{(\mrm{fb})}(\circ)= 
    \mca{L}^{(\mrm{fb})}(\circ)+\sum_{k\geq 1}(e^{iuc_{k}}-1)\mca{F}_k[L_{k}\circ L_{k}^\dagger].
\end{equation}
Therefore, $Z(u)=\tr[e^{\mca{L}_u^{\rm (fb)}\tau}(\varrho_0)]$.

Using this formulation, an explicit expression for the current average can be obtained as follows:
\begin{align}
  \eval{(-i\partial_u) Z (u)}_{u=0}
  &=\eval{-i\partial_u \tr(\varrho_{\tau,u})}_{u=0}
  \notag\\
  &=\int_0^\tau \dd{t}  \eval{(-i\partial_u) \tr(\dot{\varrho}_{t,u})}_{u=0}
  \notag\\
  &=\int_0^\tau \dd{t}  \eval{(-i\partial_u) \tr[\mca{L}_u^{(\mrm{fb})}(\varrho_{t,u})]}_{u=0}
  \notag\\
  &=\int_0^\tau \dd{t} {\sum_{k\geq 1} c_k \tr(\mca{F}_k[L_k \varrho_{t} L_k^\dagger])} - i\int_0^\tau \dd{t} \tr[\mca{L}^{\rm (fb)}\qty({\partial_u\varrho_{t,u}}|_{u=0})]
  \notag\\
  &=\int_0^\tau \dd{t}  \sum_{k\geq 1} c_k \tr(L_k \varrho_{t} L_k^\dagger),
\end{align}
where we use the facts that each feedback map $\mathcal{F}_k$ is trace-preserving and $\tr(\mca{L}^{\rm (fb)}[\circ])=0$.
Therefore,
\begin{align}
  \ev{J}=\int_0^\tau \dd{t}  \sum_{k\geq 1} c_k \tr(L_k \varrho_{t} L_k^\dagger).
\end{align}

\section{The second law of thermodynamics under feedback}
\subsection{Proof of the second law}
We first prove that the total entropy production before feedback is always nonnegative, $S(\varrho_{t+dt}^-)-S(\varrho_t)+dS^{\mathrm{env}} \ge 0$.
Assuming that $\varrho_t$ is full rank, it follows from Proposition \ref{prop:full.rank} that
\begin{align}
  S(\varrho_{t+dt}^-)-S(\varrho_t)
  &= S\!\left(\varrho_t + dt\,\mathcal{L}^{(0)}[\varrho_t]\right)
     - S(\varrho_t) \notag\\
  &= - \tr(\mathcal{L}^{(0)}[\varrho_t]\ln \varrho_t)dt + O(dt^2).
\end{align}
Here, the superoperators are defined as
\begin{align}
  \mathcal{L}^{(0)}[\circ]
  &\coloneqq -i[H,\circ] + \sum_{k\ge 1} \mathcal{D}[L_k]\circ , \\
  \mathcal{D}[L_k]\circ
  &\coloneqq L_k \circ L_k^\dagger
  - \frac{1}{2}\{L_k^\dagger L_k,\circ\} .
\end{align}
Hence,
\begin{align}
  S(\varrho_{t+dt}^-)-S(\varrho_t)+dS^{\mathrm{env}}
  = \qty{-\tr(\mathcal{L}^{(0)}[\varrho_t]\ln \varrho_t)
  + \sum_{k\ge 1} \Delta s_k \tr(L_k \varrho_t L_k^\dagger)}dt
  \eqqcolon \sigma\, dt.
\end{align}
Therefore, it suffices to show $\sigma \ge 0$.
Let $\varrho_t=\sum_n p_n(t)\dyad{n_t}$ be the spectral decomposition of $\varrho_t$, and define $w_k^{mn} \coloneqq |\mel{m_t}{L_k}{n_t}|^2 $.
The local detailed balance condition $L_k = e^{\Delta s_k/2} L_{k^*}^\dagger$ implies $w_k^{mn} = e^{\Delta s_k} w_{k^*}^{nm}$.
Using these relations, $\sigma$ can be rewritten as
\begin{align}\label{eq:setup.SL.meas}
  \sigma
  &= \sum_{k\ge 1}
     \tr(-\mathcal{D}[L_k]\varrho_t \ln \varrho_t)
     + \sum_{k\ge 1}
       \Delta s_k \tr(L_k \varrho_t L_k^\dagger)\notag \\
  &= \frac{1}{2}
     \sum_{k\ge 1}\sum_{m,n} \qty(w_k^{mn} p_n - w_{k^*}^{nm} p_m)\ln\frac{w_k^{mn} p_n}{w_{k^*}^{nm} p_m}\ge 0,
\end{align}
where the last inequality follows from the positivity of
$(x-y)\ln(x/y)$ for any $x,y\ge 0$.
This establishes the second law for the measurement step [cf.~Eq.~(\SecLawMs) in the main text].

Next, we consider the second law for the feedback step.
As shown in Proposition \ref{prop:unital}, the unitality of $\mathcal{F}_k$ implies that
\begin{align}
  S(\varrho_{t+dt,k}) - S(\varrho_{t+dt,k}^-) \ge 0 \label{eq:setup.SLfb}
\end{align}
for all $k$.
Particularly, the equality can be attained when $\mathcal{F}_k$ is a unitary map (i.e., $\mathcal{F}_k[\circ]=U_k \circ U_k^\dagger$ for some unitary operator $U_k$).
In general, the von Neumann entropy of a classical-quantum state $\varrho = \sum_k p_k \varrho_k \otimes \ket{k}\bra{k}$ can be expressed as 
\begin{align}
  S(\varrho) = H(p) + \sum_k p_k S(\varrho_k),
  \label{eq:tot.en}
\end{align}
where $H(p)=-\sum_k p_k \ln p_k$ is the Shannon entropy of the probability distribution $\{p_k\}$.
Thus, the entropy of the joint system (system + memory) after measurement can be decomposed into the memory entropy and the average entropy of the system conditioned on the measurement outcome.
Using Eq.~\eqref{eq:tot.en}, the mutual information after measurement but before feedback can be calculated as
\begin{align}
  I_{t+dt}^-
  &= S(\varrho_{t+dt}^-)+H(p)
     - S(\varrho_{t+dt,\mathrm{tot}}^-)\notag \\
  &= S(\varrho_{t+dt}^-)
     - \sum_{k\ge 0} p_k S(\varrho_{t+dt,k}^-),
\end{align}
while the mutual information after feedback is given by
\begin{align}
  I_{t+dt}
  &= S(\varrho_{t+dt})+H(p)
     - S(\varrho_{t+dt,\mathrm{tot}})\notag \\
  &= S(\varrho_{t+dt})
     - \sum_{k\ge 0} p_k S(\varrho_{t+dt,k}).
\end{align}
The quantity $I_{t+dt}^-$ is the Holevo information of the conditional state of the system immediately after measurement, whereas $I_{t+dt}$ is the corresponding Holevo information after feedback.
Using Eq.~\eqref{eq:setup.SLfb}, we obtain
\begin{align}
  0
  &\le \sum_{k\ge 0} p_k[S(\varrho_{t+dt,k})
        - S(\varrho_{t+dt,k}^-)]\notag \\
  &= S(\varrho_{t+dt})
     - S(\varrho_{t+dt}^-)
     + I_{t+dt}^- - I_{t+dt} \notag\\
     &= S(\varrho_{t+dt})
     - S(\varrho_{t+dt}^-) - dI.\label{eq:SLfb}
\end{align}
This establishes the second law for the feedback step [cf.~Eq.~(\SecLawFb) in the main text].
In particular, for unitary feedback,
$S(\varrho_{t+dt,k})=S(\varrho_{t+dt,k}^-)$ holds for all $k$, and hence $S(\varrho_{t+dt})
- S(\varrho_{t+dt}^-) -dI = 0$.

Finally, combining Eqs.~\eqref{eq:setup.SL.meas} and \eqref{eq:SLfb} yields the desired second law of thermodynamics for the total process [cf.~Eq.~(\SecRes) in the main text].

\subsubsection{Mutual information rate in the continuous-time limit}
Here we provide the detailed derivation of the mutual information rate in the continuous-time limit $dt\to 0$.

After measurement for the time interval $dt$, the composite state of the system and the memory is given by
\begin{align}
  \varrho_{t+dt,\mathrm{tot}}^- = \sum_{k\ge 0} p_k \varrho_{t+dt,k}^- \otimes \dyad{k}.
\end{align}
Here, $p_0 = \tr(M_0 \varrho_t M_0^\dagger)
  = 1 - \sum_{k\ge 1} \tr(L_k \varrho_t L_k^\dagger) dt$ and $p_k = \tr(M_k \varrho_t M_k^\dagger) = \tr(L_k \varrho_t L_k^\dagger) dt$ for $k\ge 1$.
We first evaluate the mutual information after measurement $I_{t+dt}^-$.
As shown above, $I_{t+dt}^- = S(\varrho_{t+dt}^-) - \sum_{k\ge 0} p_k S(\varrho_{t+dt,k}^-)$.
We calculate this quantity to first order in $dt$, neglecting higher-order terms.
Assuming $\varrho_t$ has full rank and applying Proposition \ref{prop:full.rank}, we have
\begin{align}
  S(\varrho_{t+dt}^-)&=S(\varrho_t+\mca{L}^{(0)}[\varrho_t]dt) \notag  \\
  &=S(\varrho_t)-\tr{[-i[H,\varrho_t]+ \sum_{k\geq 1} (\mca{D}[L_k]\varrho_t )]\ln \varrho_t}dt \notag\\
  &=S(\varrho_t)-\sum_{k\geq 1}\tr(L_k \varrho_t L_k^\dagger \ln \varrho_t)dt
  +\sum_{k\geq 1}\tr(\{L_k^\dagger L_k ,\varrho_t \}/2\ln \varrho_t)dt.
\end{align}
Noting that $\varrho_{t+dt,0}^- =M_0 \varrho_t M_0^\dagger/p_0=\qty[\varrho_t+(-i[H,\varrho_t]- \sum_{k\geq 1} \{L_k^\dagger L_k ,\varrho_t \}/2)dt]/p_0$ and $\varrho_{t+dt,k}^-=M_k \varrho_t M_k^\dagger /p_k=L_k \varrho_t L_k^\dagger dt /p_k$, we obtain
\begin{align}
  p_0 S(\varrho_{t+dt,0}^-)&=
  -\tr[M_0 \varrho_t M_0^\dagger \ln(M_0 \varrho_t M_0^\dagger)]
  +p_0 \ln p_0
  \notag\\
  &=S(\varrho_t)
  -\tr[(-i[H,\varrho_t]- \sum_{k\geq 1} \{L_k^\dagger L_k ,\varrho_t \}/2)\ln \varrho_t]dt
  -\tr[-i[H,\varrho_t]- \sum_{k\geq 1} \{L_k^\dagger L_k ,\varrho_t \}/2]dt \notag
  \notag\\
  &+(1-\sum_{k\geq 1} \tr[L_k \varrho_t L_k^\dagger ]dt)\ln (1-\sum_{k\geq 1} \tr[L_k \varrho_t L_k^\dagger ]dt)
  \notag\\
  &=S(\varrho_t)
  +\sum_{k\geq 1}\tr[ (\{L_k^\dagger L_k ,\varrho_t \}/2)\ln \varrho_t]dt
  +\sum_{k\geq 1} \tr(L_k \varrho_t L_k^\dagger)dt
  -\sum_{k\geq 1} \tr(L_k \varrho_t L_k^\dagger)dt
  \notag\\
  &=S(\varrho_t)
  +\sum_{k\geq 1}\tr[ (\{L_k^\dagger L_k ,\varrho_t \}/2)\ln \varrho_t]dt,
\end{align}
and
\begin{align}
  p_k S(\varrho_{t+dt,k}^-)&=
  -\tr[M_k \varrho_t M_k^\dagger \ln(M_k \varrho_t M_k^\dagger)]
  +p_k \ln p_k
  \notag\\
  &=-\tr[L_k \varrho_t L_k^\dagger dt \ln (L_k \varrho_t L_k^\dagger dt)]
  +\tr(L_k \varrho_t L_k^\dagger dt)\ln \tr(L_k \varrho_t L_k^\dagger dt)
  \notag\\
  &=-\tr[L_k \varrho_t L_k^\dagger \ln (L_k \varrho_t L_k^\dagger)]dt
  -\tr(L_k \varrho_t L_k^\dagger) dt \ln dt +\qty[\tr(L_k \varrho_t L_k^\dagger)\ln \tr(L_k \varrho_t L_k^\dagger)]dt
  +\tr(L_k \varrho_t L_k^\dagger)dt\ln dt
  \notag\\
  &=-\tr[L_k \varrho_t L_k^\dagger \ln (L_k \varrho_t L_k^\dagger)]dt 
+[\tr(L_k \varrho_t L_k^\dagger)\ln \tr(L_k \varrho_t L_k^\dagger)]dt.
\end{align}
Combining these, we obtain
\begin{align}
  I_{t+dt}^- &= S(\varrho_{t+dt}^-)-\sum_{k\geq 0} p_k S(\varrho_{t+dt,k}^-)
  \notag\\
  &= S(\varrho_t)-\sum_{k\geq 1}\tr(L_k \varrho_t L_k^\dagger \ln \varrho_t)dt
  +\sum_{k\geq 1}\tr[(\{L_k^\dagger L_k ,\varrho_t \}/2) \ln \varrho_t]dt \notag
  \\
  &-S(\varrho_t)
  -\sum_{k\geq 1} \tr[(\{L_k^\dagger L_k ,\varrho_t \}/2)\ln \varrho_t]dt \notag
  \\
  &+ \sum_{k\geq 1}\qty
  {\tr[L_k \varrho_t L_k^\dagger \ln (L_k \varrho_t L_k^\dagger)]
  - \tr(L_k \varrho_t L_k^\dagger)\ln \tr(L_k \varrho_t L_k^\dagger)}dt
  \notag\\
  &= \sum_{k\geq 1}\qty{-\tr(L_k \varrho_t L_k^\dagger \ln \varrho_t)
  +\tr[L_k \varrho_t L_k^\dagger \ln (L_k \varrho_t L_k^\dagger)]
  -\tr(L_k \varrho_t L_k^\dagger)\ln \tr(L_k \varrho_t L_k^\dagger)}dt.
\end{align}
Using the quantum relative entropy, $I_{t+dt}^-$ can be expressed as follows:
\begin{align}
  I_{t+dt}^- 
  &= \sum_{k\geq 1}\qty{
  \tr(L_k \varrho_t L_k^\dagger \ln \frac{L_k \varrho_t L_k^\dagger}{\tr[L_k \varrho_t L_k^\dagger]})
  -\tr(L_k \varrho_t L_k^\dagger \ln \varrho_t)}dt
  \notag\\
  &= \sum_{k\geq 1}\tr(L_k \varrho_t L_k^\dagger)
  \qty{\tr(\frac{L_k \varrho_t L_k^\dagger}{\tr[L_k \varrho_t L_k^\dagger]}  \ln \frac{L_k \varrho_t L_k^\dagger}{\tr[L_k \varrho_t L_k^\dagger]} )
  -\tr(\frac{L_k \varrho_t L_k^\dagger}{\tr[L_k \varrho_t L_k^\dagger]} \ln \varrho_t)}dt\notag
  \\
  &= \sum_{k\geq 1}\tr(L_k \varrho_t L_k^\dagger)
  D\qty(\frac{L_k \varrho_t L_k^\dagger}{\tr[L_k \varrho_t L_k^\dagger]} \|\varrho_t)dt\notag
  \\
  &=\sum_{k\geq 1}p_k
  D\qty(\frac{L_k \varrho_t L_k^\dagger}{\tr[L_k \varrho_t L_k^\dagger]} \|\varrho_t)
  \geq 0.
\end{align}

Next, we evaluate the mutual information $I_{t+dt}$ after feedback.
To first order in $dt$, the von Neumann entropy can be similarly calculated as
\begin{align}
  S(\varrho_{t+dt})&=S(\varrho_t+\mca{L}^{\mrm{(fb)}}[\varrho_t]dt)
  \notag\\
  &=S(\varrho_t)-\tr[\mca{L}^{\mrm{(fb)}}[\varrho_t]\ln \varrho_t]dt
  \notag\\
  &=S(\varrho_t)-\sum_{k\geq 1}\tr(\mca{F}_k[L_k \varrho_t L_k^\dagger] \ln \varrho_t)dt
  +\sum_{k\geq 1}\tr[(\{L_k^\dagger L_k ,\varrho_t \}/2)\ln \varrho_t]dt.
\end{align}
Note that $\varrho_{t+dt,0}=\varrho_{t+dt,0}^-$ and $\varrho_{t+dt,k}=\mca{F}_k [\varrho_{t+dt,k}^-]=\mca{F}_k[L_k \varrho_t L_k^\dagger] dt /p_k$.
Using the fact that each $\mca{F}_k$ is trace-preserving, we obtain
\begin{align}
  p_k S(\varrho_{t+dt,k})&=
  -\tr(\mca{F}_k[M_k \varrho_t M_k^\dagger] \ln \mca{F}_k[M_k \varrho_t M_k^\dagger])
  +p_k \ln p_k
  \notag\\
  &=-\tr(\mca{F}_k[L_k \varrho_t L_k^\dagger dt] \ln \mca{F}_k[L_k \varrho_t L_k^\dagger dt])
  +\tr(L_k \varrho_t L_k^\dagger dt)\ln \tr(L_k \varrho_t L_k^\dagger dt)
  \notag\\
  &=-\tr(\mca{F}_k[L_k \varrho_t L_k^\dagger] \ln \mca{F}_k[L_k \varrho_t L_k^\dagger])dt
  -\tr(L_k \varrho_t L_k^\dagger) dt \ln dt \notag
  \\
  &+ [\tr(L_k \varrho_t L_k^\dagger)\ln \tr(L_k \varrho_t L_k^\dagger)]dt
  +\tr(L_k \varrho_t L_k^\dagger) dt\ln dt
  \notag\\
  &=\qty{-\tr(\mca{F}_k[L_k \varrho_t L_k^\dagger] \ln \mca{F}_k[L_k \varrho_t L_k^\dagger])
+ \tr(L_k \varrho_t L_k^\dagger)\ln \tr(L_k \varrho_t L_k^\dagger)}dt.
\end{align}
Combining these, we obtain
\begin{align}
  I_{t+dt} &= S(\varrho_{t+dt})-\sum_{k\geq 0} p_k S(\varrho_{t+dt,k})
  \notag\\
  &= S(\varrho_t)-\sum_{k\geq 1}\tr(\mca{F}_k[L_k \varrho_t L_k^\dagger] \ln \varrho_t)dt
  +\sum_{k\geq 1}\tr[(\{L_k^\dagger L_k ,\varrho_t \}/2)\ln \varrho_t]dt \notag
  \\
  &-S(\varrho_t)
  -\sum_{k\geq 1} \tr[(\{L_k^\dagger L_k ,\varrho_t \}/2)\ln \varrho_t]dt \notag
  \\
  &+ \sum_{k\geq 1}\qty
  {\tr(\mca{F}_k[L_k \varrho_t L_k^\dagger] \ln \mca{F}_k[L_k \varrho_t L_k^\dagger])
  - \tr(L_k \varrho_t L_k^\dagger)\ln \tr(L_k \varrho_t L_k^\dagger)}dt
  \notag\\
  &= \sum_{k\geq 1}\qty{-\tr(\mca{F}_k[L_k \varrho_t L_k^\dagger] \ln \varrho_t)
  +\tr(\mca{F}_k[L_k \varrho_t L_k^\dagger] \ln \mca{F}_k[L_k \varrho_t L_k^\dagger])
  -\tr(L_k \varrho_t L_k^\dagger)\ln \tr(L_k \varrho_t L_k^\dagger)}dt.
\end{align}
This can be further expressed in terms of the quantum relative entropy as
\begin{align}
  I_{t+dt}
  &= \sum_{k\geq 1}\qty{
  \tr(\mca{F}_k[L_k \varrho_t L_k^\dagger] \ln \frac{\mca{F}_k[L_k \varrho_t L_k^\dagger]}{\tr[L_k \varrho_t L_k^\dagger]})
  -\tr(\mca{F}_k[L_k \varrho_t L_k^\dagger] \ln \varrho_t)}dt\notag
  \\
  &= \sum_{k\geq 1}\tr(L_k \varrho_t L_k^\dagger)
  \qty{\tr(\frac{\mca{F}_k[L_k \varrho_t L_k^\dagger]}{\tr[L_k \varrho_t L_k^\dagger]}  \ln \frac{\mca{F}_k[L_k \varrho_t L_k^\dagger]}{\tr[L_k \varrho_t L_k^\dagger]} )
  -\tr(\frac{\mca{F}_k[L_k \varrho_t L_k^\dagger]}{\tr[L_k \varrho_t L_k^\dagger]} \ln \varrho_t)}dt
  \notag\\
  &= \sum_{k\geq 1}\tr(L_k \varrho_t L_k^\dagger)
  D\qty(\frac{\mca{F}_k[L_k \varrho_t L_k^\dagger]}{\tr[L_k \varrho_t L_k^\dagger]} \|\varrho_t)dt
  \notag\\
  &=\sum_{k\geq 1}p_k
  D\qty(\frac{\mca{F}_k[L_k \varrho_t L_k^\dagger]}{\tr[L_k \varrho_t L_k^\dagger]} \|\varrho_t)
  \geq 0.
\end{align}
Consequently, the mutual information rate $\dot I=\lim_{dt\to 0}(I_{t+dt}-I_{t+dt}^-)/dt$ can be calculated as
\begin{equation}
	\dot I=\sum_{k\ge 1}\tr(L_k\varrho_tL_k^\dagger)\qty{D\qty(\mca{F}_k[\varrho_{t,k}'] \| \varrho_t) - D\qty(\varrho_{t,k}' \| \varrho_t)},
\end{equation}
where $\varrho_{t,k}'\coloneqq L_k \varrho_t L_k^\dagger/\tr(L_k \varrho_t L_k^\dagger)$.

\subsection{Comparison with previous formulations of the second law}\label{app:sl.prx}
Several versions of the second law that incorporate measurement and feedback have been proposed in the literature. Among them, Ref.~\cite{Strasberg.2017.PRX} considers a setting closely related to ours. In order to clarify the position of the second law we proved, we compare our formulation with that of Ref.~\cite{Strasberg.2017.PRX}.

The setting in Sec.~IV.E of Ref.~\cite{Strasberg.2017.PRX} is most relevant to our framework. There, a unit $U$ is introduced as a memory that records the measurement outcomes. The measurement process is described by a superoperator $\mathcal{L}$ acting on the composite system $S+U$ for a finite time, which encompasses our measurement scheme.
The feedback process is implemented as a unitary evolution of the composite system comprising the target system and its environment, conditioned on the measurement outcome. In our setup, feedback is implemented through unital channels, which may involve heat dissipation. By the Stinespring dilation theorem, any completely positive trace-preserving (CPTP) map can be represented as a unitary evolution on an extended Hilbert space with a suitably initialized environment. Hence, the feedback process considered in Ref.~\cite{Strasberg.2017.PRX} includes our setting as a special case. For a direct comparison, we specialize to feedback without heat dissipation, for which the previous formulation of the feedback second law reduces to ours.

We first rewrite the second law of the measurement step in Ref.~\cite{Strasberg.2017.PRX} [Eq.~(71) therein] using our notations. 
The corresponding thermodynamic quantities are given by
\begin{align}
  \Delta S^{\mathrm{ms}}_S &= S(\varrho_{t+dt}^-) - S(\varrho_t),
  \\
  \Delta S^{\mathrm{ms}}_U &= H(p) - 0,
  \\
  I^{\mathrm{ms}}_{S:U}
  &= I_{t+dt}^-
  = S(\varrho_{t+dt}^-)
    - \sum_{k\ge 0} p_k S(\varrho_{t+dt,k}^-),
  \\
  -\beta Q^{\mathrm{ms}} &= dS^{\mathrm{env}}.
\end{align}
Then, the second law for the measurement step formulated in Ref.~\cite{Strasberg.2017.PRX} reads
\begin{equation}
    \Sigma^{\mathrm{ms}}
  = \Delta S^{\mathrm{ms}}_S
     + \Delta S^{\mathrm{ms}}_U
     - I^{\mathrm{ms}}_{S:U}
     - \beta Q^{\mathrm{ms}}\ge 0.
\end{equation}
Using the expressions above, $\Sigma^{\mathrm{ms}}$ can be explicitly calculated as
\begin{align}
  \Sigma^{\mathrm{ms}}
  &= \Delta S^{\mathrm{ms}}_S
     + \Delta S^{\mathrm{ms}}_U
     - I^{\mathrm{ms}}_{S:U}
     - \beta Q^{\mathrm{ms}}
  \notag\\
  &= S(\varrho_{t+dt}^-) - S(\varrho_t)
     + H(p)
     - S(\varrho_{t+dt}^-)
     + \sum_{k\ge 0} p_k S(\varrho_{t+dt,k}^-)
     + dS^{\mathrm{env}}
  \notag\\
  &= - S(\varrho_t)
     + H(p)
     + \sum_{k\ge 0} p_k S(\varrho_{t+dt,k}^-)
     + dS^{\mathrm{env}}.\label{eq:ms.ent.ref71}
\end{align}
By applying the inequality on the entropy of a mixture of quantum states \cite{Nielsen.2000}, we have
\begin{align}
  S(\varrho_{t+dt}^-)=S\qty( \sum_{k\ge 0} p_k \varrho_{t+dt,k}^- )
  \le
  H(p) + \sum_{k\ge 0} p_k S(\varrho_{t+dt,k}^-).\label{eq:mix.ent.ine}
\end{align}
Combining Eqs.~\eqref{eq:ms.ent.ref71} and \eqref{eq:mix.ent.ine} yields the following inequality:
\begin{align}
  \Sigma^{\mathrm{ms}}
  \ge
  S(\varrho_{t+dt}^-) - S(\varrho_t)
  + dS^{\mathrm{env}}\ge 0.
\end{align}
Hence, our second law for the measurement step is tighter than the extant one.

Next, we rewrite the second law of the feedback step in Ref.~\cite{Strasberg.2017.PRX} [Eq.~(75) therein] for our setting. Using our notations, the corresponding thermodynamic quantities are given by
\begin{align}
  \Delta S^{\mathrm{fb}}_S
  &= S(\varrho_{t+dt}) - S(\varrho_{t+dt}^-),
  \\
  \Delta S^{\mathrm{fb}}_U &= 0,
  \\
  I^{\mathrm{fb}}_{S:U}
  &= I_{t+dt}
  = S(\varrho_{t+dt})
    - \sum_{k\ge 0} p_k S(\varrho_{t+dt,k}),
  \\
  -\beta Q^{\mathrm{fb}} &= 0.
\end{align}
Then, the second law for the feedback step reads
\begin{equation}
    \Sigma^{\mathrm{fb}}
  = \Delta S^{\mathrm{fb}}_S
     + \Delta S^{\mathrm{fb}}_U
     + I^{\mathrm{ms}}_{S:U}
     - \beta Q^{\mathrm{fb}}\ge I^{\mathrm{fb}}_{S:U}\ge 0.
\end{equation}
This can be further evaluated as
\begin{align}
  \Sigma^{\mathrm{fb}}
  &= S(\varrho_{t+dt}) - S(\varrho_{t+dt}^-)
     + S(\varrho_{t+dt}^-)
    - \sum_{k\ge 0} p_k S(\varrho_{t+dt,k}^-)
  \notag\\
  &= S(\varrho_{t+dt})
     - \sum_{k\ge 0} p_k S(\varrho_{t+dt,k}^-).
\end{align}
Therefore, the refined second law $\Sigma^{\rm fb}-I_{S:U}^{\rm fb}\ge 0$ can be expressed as
\begin{align}
  \Sigma^{\mathrm{fb}} - I^{\mathrm{fb}}_{S:U}
  &= S(\varrho_{t+dt})
     - \sum_{k\ge 0} p_k S(\varrho_{t+dt,k}^-)
     - S(\varrho_{t+dt})
     + \sum_{k\ge 0} p_k S(\varrho_{t+dt,k})
  \\
  &= \sum_{k\ge 0} p_k
     \qty[
       S(\varrho_{t+dt,k})
       - S(\varrho_{t+dt,k}^-)
     ]\notag\\
    &= S(\varrho_{t+dt})-S(\varrho_{t+dt}^-)-dI \ge 0,
\end{align}
which is identical to our second law for the feedback step.

Taking both the measurement and feedback processes into account, we find that our formulation of the second law yields a tighter bound. By comparison, the formulation in Ref.~\cite{Strasberg.2017.PRX} allows for more general measurement dynamics, but at the expense of tightness.

\section{Detailed derivation of the central result (\FirRes)}\label{app:derivation}
Here, we provide a complete derivation of our central result [Eq.~(\FirRes)], following the approach of Refs.~\cite{Vu.2025.PRXQ,Honma.2026.PRA}, which proceeds in three steps:
\begin{enumerate}
	\item[(1)] Generalized quantum Cram{\'e}r-Rao inequality: In Appendix \ref{app:gen.CR.ine}, we derive the following generalized quantum Cram{\'e}r-Rao inequality for Markovian feedback dynamics:
\begin{equation}\label{eq:gen.CR.ine}
	\frac{\Var[J]_\theta}{(\partial_\theta\ev{J}_\theta)^2}\ge\frac{1}{\mca{I}_{\theta}}.
\end{equation}
	
	\item[(2)] Calculation of quantum Fisher information: In Appendix \ref{app:quantum.Fisher}, we show how the quantum Fisher information $\mca{I}_{\theta}$ can be explicitly computed for Markovian feedback dynamics.
	
	\item[(3)] Evaluation under perturbation: In Appendix \ref{app:proof.qtkur}, we evaluate $\mca{I}_0$ and $\partial_\theta\ev{J}_\theta|_{\theta=0}$ for the perturbed dynamics.
\end{enumerate}

\subsection{Generalized quantum Cram{\'e}r-Rao inequality}\label{app:gen.CR.ine}
For convenience, we consider a general Markovian feedback dynamics parameterized by a scalar parameter $\theta$,
\begin{equation}\label{eq:gen.mas}
	\dot\varrho_{t,\theta}=-i[H_{\theta},\varrho_{t,\theta}]
  +\sum_{k\ge 1}\qty(\mca{F}_k[L_{k,\theta}\varrho_{t,\theta}L_{k,\theta}^\dagger]-\frac{1}{2}\{L_{k,\theta}^\dagger L_{k,\theta},\varrho_{t,\theta}\}),
\end{equation}
where the initial state is independent of $\theta$, i.e., $\varrho_{0,\theta}=\varrho_0$. 
Feedback channels $\{\mca{F}_k\}$ are CPTP maps independent of $\theta$.
For simplicity, we consider an initial pure state $\varrho_0=\dyad{\psi_0}$, as any mixed initial state can be purified using an ancillary system, and the same analysis can be applied straightforwardly.
We discretize the total evolution time as $\tau=N \Delta t$, where $N$ is a positive integer, and the continuous-time limit can be achieved by taking the limit $N \rightarrow \infty$.
In the short-time limit, the master equation \eqref{eq:gen.mas} can be expressed in the form of the Kraus representation as
\begin{equation}\label{eq:Kraus.rep}
	\varrho_{t+\Delta t,\theta}=M_{0,\theta}\varrho_{t,\theta}M_{0,\theta}^\dagger
  +\sum_{k\ge 1}\mca{F}_k[M_{k,\theta}\varrho_{t,\theta}M_{k,\theta}^\dagger],
\end{equation}
where the Kraus operators $\{M_{k,\theta}\}$ satisfy the completeness condition $\sum_{k\ge 0}M_{k,\theta}^\dagger M_{k,\theta}=\mbb{1}_S$.
Here, $\mbb{1}_\alpha$ denotes the identity operator on the Hilbert space of system $\alpha$.
It is worth noting that the unraveling is not unique.
In the case of quantum jump unraveling, the Kraus operators are given by
\begin{equation}
	M_{0,\theta}=\mbb{1}_S-iH_{\rm eff,\theta}\Delta t,~M_{k,\theta}=L_{k,\theta}\sqrt{\Delta t},
\end{equation}
where the effective Hamiltonian is defined as $H_{\rm eff,\theta}\coloneqq H_\theta-(i/2)\sum_{k\ge 1}L_{k,\theta}^\dagger L_{k,\theta}$.
The Kraus representation in Eq.~\eqref{eq:Kraus.rep} enables us to unravel the deterministic feedback dynamics into an ensemble of stochastic trajectories $\{\Gamma = (k_1, \dots, k_N)\}$, where each operator $M_{k_i,\theta}$ accounts for the evolution during the time interval $[(i-1)\Delta t, i\Delta t]$.
We consider the Stinespring representation for each feedback channel $\mca{F}_k$.
For each time interval $[(i-1)\Delta t, i\Delta t]$, we prepare a sufficiently large environment initialized in state $\ket{0}_{i}$.
The initial state of the composite system of the target system and the environment is $\ket{\psi_0}\otimes \ket{0}_R$, where $\ket{0}_R\coloneqq \ket{0}_{N}\otimes \dots \otimes \ket{0}_1$.
As the Stinespring representation of $\{\mca{F}_{k}\}$, one can always find unitary operators $\{V_{k}\}$ such that $\mca{F}_{k_i} [\circ]=\tr_{R_i}[V_{k_i} \circ \otimes\ket{0}_i \bra{0}_i V_{k_i}^\dagger]$.
Additionally, for simplicity of notation, we define the identity map $\mca{F}_0[\circ]=\circ$,
which can be represented with $V_0=\mbb{1}_S\otimes \mbb{1}_R$.
The jump events and the final pure state of the joint system and environment for all trajectories can be encoded into the following extended state:
\begin{equation}
	\ket{\Psi_{\tau,\theta}}\coloneqq\sum_{\Gamma}V_{k_N} M_{k_{N},\theta}\dots V_{k_1}M_{k_1,\theta}\ket{\psi_0}\otimes\ket{0}_R \otimes \ket{\Gamma},
\end{equation}
where the virtual states $\ket{\Gamma}\coloneqq\ket{k_{N},\dots,k_1}$ record jump information and form an orthonormal basis.
For each trajectory $\Gamma$, the current $J$ can be calculated as
\begin{equation}
	J=J(\Gamma)\coloneqq\sum_{i=1}^{N}c_{k_i},
\end{equation}
where we set $c_0=0$.
We then define the self-adjoint current operator:
\begin{equation}
	\msf{J}\coloneqq\mbb{1}_S\otimes \mbb{1}_R\otimes \sum_{\Gamma}J(\Gamma)\dyad{\Gamma},
\end{equation}
so that the average and variance of the current can be expressed as
\begin{align}
	\ev{J}_\theta&=\mel{\Psi_{\tau,\theta}}{\msf{J}}{\Psi_{\tau,\theta}},\\
	\Var[J]_\theta&=\mel{\Psi_{\tau,\theta}}{(\msf{J}-\ev{J}_\theta)^2}{\Psi_{\tau,\theta}}.
\end{align}
We define the density operator $\Psi_{\tau,\theta}\coloneqq\dyad{\Psi_{\tau,\theta}}$ for the pure state $\ket{\Psi_{\tau,\theta}}$ and introduce the Symmetric Logarithmic Derivative (SLD) operator $\msf{L}_\theta$ via
\begin{equation}
	\partial_\theta{\Psi_{\tau,\theta}}=\frac{1}{2}\qty( \msf{L}_\theta{\Psi_{\tau,\theta}}+{\Psi_{\tau,\theta}}\msf{L}_\theta ).
\end{equation}
The quantum Fisher information is then defined using the SLD operator as
\begin{equation}\label{eq:qFI.def}
	\mca{I}_{\theta}\coloneqq\tr(\msf{L}_\theta^2{\Psi_{\tau,\theta}})=\mel{\Psi_{\tau,\theta}}{\msf{L}_\theta^2}{\Psi_{\tau,\theta}}.
\end{equation}
Applying the Cauchy-Schwarz inequality, we obtain
\begin{align}
	|\partial_\theta\ev{J}_\theta|&=|\tr(\msf{J}\partial_\theta{\Psi_{\tau,\theta}})|\notag\\
	&=|\tr{(\msf{J}-\ev{J}_\theta)\partial_\theta {\Psi_{\tau,\theta}}}|\notag\\
	&=\frac{1}{2}|\mel{\Psi_{\tau,\theta}}{(\msf{J}-\ev{J}_\theta)\msf{L}_\theta+\msf{L}_\theta(\msf{J}-\ev{J}_\theta)}{\Psi_{\tau,\theta}}|\notag\\
	&\le \sqrt{\mca{I}_{\theta}\Var[J]_\theta},
\end{align}
which leads directly to the generalized Cram{\'e}r-Rao inequality:
\begin{equation}\label{eq:app.Cramer-Rao1}
	\frac{\Var[J]_\theta}{(\partial_\theta\ev{J}_\theta)^2}\ge\frac{1}{\mca{I}_{\theta}}.
\end{equation}
It is worth emphasizing that this inequality holds for arbitrary trajectory observables, extending beyond simple counting-type observables.

\subsection{Calculation of quantum Fisher information}\label{app:quantum.Fisher}
Next, we describe the calculation of quantum Fisher information $\mca{I}_{\theta}$ for GKSL dynamics, following the approach in Ref.~\cite{Gammelmark.2014.PRL}.
Taking the derivative of pure state $\Psi_{\tau,\theta}$ with respect to parameter $\theta$ yields $\partial_\theta\Psi_{\tau,\theta} =\dyad{\partial_\theta \Psi_{\tau,\theta}}{\Psi_{\tau,\theta}}+\dyad{\Psi_{\tau,\theta}}{\partial_\theta \Psi_{\tau,\theta}}$.
From this relation and the normalization condition $\braket{\Psi_{\tau,\theta}}{\Psi_{\tau,\theta}}=1$, we can verify that the SLD operator can be expressed as
\begin{align}
  \msf{L}_\theta= 2(\dyad{\partial_\theta \Psi_{\tau,\theta}}{\Psi_{\tau,\theta}}+\dyad{\Psi_{\tau,\theta}}{\partial_\theta \Psi_{\tau,\theta}}).
\end{align}
Substituting this expression into Eq.~\eqref{eq:qFI.def}, we obtain
\begin{align}
  \mca{I}_{\theta}&=4\qty( \braket{\partial_\theta \Psi_{\tau,\theta}}{\partial_\theta \Psi_{\tau,\theta}} - |\braket{\Psi_{\tau,\theta}}{\partial_\theta \Psi_{\tau,\theta}}|^2 ).
\end{align}
Here, we use the fact that $\braket{\partial_\theta \Psi_{\tau,\theta}}{\Psi_{\tau,\theta}}$ is a pure imaginary number, which can be derived from the identity $0=\partial_\theta\braket{\Psi_{\tau,\theta}}=\braket{\partial_\theta \Psi_{\tau,\theta}}{\Psi_{\tau,\theta}}+\braket{\Psi_{\tau,\theta}}{\partial_\theta \Psi_{\tau,\theta}}$.
Notably, the quantum Fisher information can be equivalently written as
\begin{align}
  \mca{I}_{\theta}&=4\partial_{\theta_1\theta_2}^2(\ln|\braket{\Psi_{\tau,\theta_1}}{\Psi_{\tau,\theta_2}}|)|_{\theta_1=\theta_2=\theta}.\label{eq:qFI.equiv}
\end{align}
To verify this, recall that for any complex function $\phi(x)$,
\begin{align}
  \partial_x\ln |\phi(x)|=\Re\qty[\frac{\partial_x \phi(x)}{\phi(x)}].
\end{align}
Then,
\begin{align}
  &4\eval{\partial_{\theta_1\theta_2}^2 (\ln|\braket{\Psi_{\tau,\theta_1}}{\Psi_{\tau,\theta_2}}|)}_{\theta_1=\theta_2=\theta}\notag\\
  &=4\eval{\partial_{\theta_1} \Re 
  \qty(\frac{\braket{\Psi_{\tau,\theta_1}}{\partial_{\theta_2} \Psi_{\tau,\theta_2}}}{\braket{\Psi_{\tau,\theta_1}}{\Psi_{\tau,\theta_2}}})}_{\theta_1=\theta_2=\theta}\notag\\
  &=4\qty(\braket{\partial_{\theta}\Psi_{\tau,\theta}}{\partial_{\theta} \Psi_{\tau,\theta}}-|\braket{\Psi_{\tau,\theta}}{\partial_\theta \Psi_{\tau,\theta}}|^2)\notag\\
  &=\mca{I}_{\theta}.
\end{align}
Now, note that $|\braket{\Psi_{\tau,\theta_1}}{\Psi_{\tau,\theta_2}}|=|\tr(\dyad{\Psi_{\tau,\theta_1}}{\Psi_{\tau,\theta_2}})|$ and
\begin{align}
  \tr_{A,R,\Gamma}\qty(\dyad{\Psi_{\tau,\theta_1}}{\Psi_{\tau,\theta_2}})
  &=\tr_{A,R}(\sum_{\Gamma}V_{k_N} M_{k_{N},\theta_1}\dots V_{k_1}M_{k_1,\theta_1}\dyad{\psi_0}\otimes\ket{0}_R \bra{0}_R
  M_{k_1,\theta_2}^\dagger V_{k_1}^\dagger \dots M_{k_{N},\theta_2}^\dagger V_{k_N}^\dagger )
  \notag\\
  &=\sum_{\Gamma}\mca{F}_{k_N}[ M_{k_{N},\theta_1}\dots \mca{F}_{k_1}[M_{k_1,\theta_1}\tr_{A}(\dyad{\psi_0})
  M_{k_1,\theta_2}^\dagger] \dots M_{k_{N},\theta_2}^\dagger ]
  \notag\\
  &=\sum_{\Gamma}\mca{F}_{k_N}[ M_{k_{N},\theta_1}\dots \mca{F}_{k_1}[M_{k_1,\theta_1}\varrho_0
  M_{k_1,\theta_2}^\dagger] \dots M_{k_{N},\theta_2}^\dagger ]
  \notag\\
  &\eqqcolon \varrho_{\tau,\vb*{\theta}},
\end{align}
where $\vb*{\theta}=(\theta_1,\theta_2)$ and $A$ denotes an ancillary system introduced to purify the initial mixed state $\varrho_0$.
By definition, the time evolution of the operator $\varrho_{t,\vb*{\theta}}$ can be described as
\begin{equation}\label{eq:two.side.op.evol}
	\varrho_{t+\Delta t,\vb*{\theta}}=\sum_{k\ge 0}\mca{F}_k[M_{k,\theta_1}\varrho_{t,\vb*{\theta}}M_{k,\theta_2}^\dagger].
\end{equation}
Taking the limit $\Delta t\to 0$ of Eq.~\eqref{eq:two.side.op.evol} yields the differential equation that describes the time evolution of the two-sided operator $\varrho_{t,\vb*{\theta}}$,
\begin{align}\label{eq:two.side.master.eq}
	\dot\varrho_{t,\vb*{\theta}}&=-i(H_{\theta_1}\varrho_{t,\vb*{\theta}}-\varrho_{t,\vb*{\theta}}H_{\theta_2})
  +\sum_{k\ge 1}\qty(\mca{F}_k[L_{k,\theta_1} \varrho_{t,\vb*{\theta}} L_{k,\theta_2}^\dagger]
  -\frac12 \{L_{k,\theta_1}^\dagger L_{k,\theta_1}\varrho_{t,\vb*{\theta}} + \varrho_{t,\vb*{\theta}}L_{k,\theta_2}^\dagger L_{k,\theta_2}\})
  \notag\\
	&\coloneqq\mca{L}_{t,\vb*{\theta}}(\varrho_{t,\vb*{\theta}}),
\end{align}
with the initial condition $\varrho_{0,\vb*{\theta}}=\varrho_0$.
Consequently, the quantum Fisher information can be analytically calculated as 
\begin{equation}\label{eq:qFI.gen.form}
    \mca{I}_{\theta}=4\eval{\partial_{\theta_1\theta_2}^2\qty(\ln\qty|\tr\varrho_{\tau,\vb*{\theta}}|)}_{\theta_1=\theta_2=\theta},
\end{equation}
where $\varrho_{\tau,\vb*{\theta}}=\vec{\mca{T}}e^{\int_0^\tau\dd{t}\mca{L}_{t,\vb*{\theta}}}(\varrho_0)$ and $\vec{\mca{T}}$ denotes the time-ordering operator.

\subsection{Evaluation under perturbation}\label{app:proof.qtkur}
From $S(\varrho_{t+dt})-S(\varrho_{t+dt}^-)-dI\geq 0$, it follows that $S(\varrho_{t+dt})-S(\varrho_{t})+dS^{\mrm{env}}-dI\geq S(\varrho_{t+dt}^-)-S(\varrho_{t})+dS^{\mrm{env}}$; therefore, $\dot{\Sigma}=\dot{S}^{\mrm{tot}}-\dot{I}\geq -\tr(\mca{L}^{(0)}[\varrho_t]\ln \varrho_t)+\sum_{k\geq 1}\tr(L_k\varrho_t L_k^\dagger)\Delta s_k \eqqcolon \sigma$.
Using the spectral decomposition $\varrho_t = \sum_n p_n(t)\dyad{n_t}$, we define $w^{mn}_{k}(t)\coloneqq|\mel{m_t}{L_{k}}{n_t}|^2$ and
\begin{align}
	\sigma^{mn}_k & \coloneqq  (w^{mn}_{k} p_n-w^{nm}_{k^*} p_m)\ln \frac{w^{mn}_{k} p_n}{w^{nm}_{k^*} p_m},
  \\
	a^{mn}_k &\coloneqq w^{mn}_{k} p_n+w^{nm}_{k^*} p_m.
\end{align}
Noting the relation $w_{k}^{mn} = e^{\Delta s_{k}} w_{k^*}^{nm}$ and performing algebraic manipulations, we can show that $\sigma=(1/2)\sum_{k\ge 1,m,n}\sigma^{mn}_k$ and $\dot{A}=(1/2)\sum_{k\ge 1,m,n}a^{mn}_k$ \cite{Vu.2023.PRX}.

Next, we consider the generalized quantum Cram{\'e}r-Rao bound for the following perturbed dynamics:
\begin{align}
	H_\theta=H~\text{and}~L_{k, \theta} = \sqrt{1 + \ell_{k}(t)\theta}L_{k},
\end{align}
where
\begin{equation}
	\ell_k(t)= \frac{\tr(L_k \varrho_t L_k^\dagger) - \tr(L_{k^*} \varrho_t L_{k^*}^\dagger)}{\tr(L_k \varrho_t L_k^\dagger) + \tr(L_{k^*} \varrho_t L_{k^*}^\dagger)}.
\end{equation}
The density matrix $\varrho_{t,\theta}$ of the perturbed dynamics is governed by
\begin{align}
  \dot{\varrho}_{t,\theta} &=-i[H,\varrho_{t,\theta}]
  +\sum_{k\geq 1}(1 + \ell_{k}(t)\theta)\qty(\mca{F}_k[L_k\varrho_{t,\theta} L_k^\dagger]
  -\frac{1}{2}\{L_k^\dagger L_k,\varrho_{t,\theta} \})
  \notag\\
  &\eqqcolon\mca{L}_{\theta}[\varrho_{t,\theta} ].
\end{align}
The quantum Cram{\'e}r-Rao inequality for $\theta=0$ reads 
\begin{align}\label{eq:qCRcound}
  \frac{\Var[J]}{(\partial_\theta \ev{J}_\theta |_{\theta=0})^2}\geq \frac{1}{\mca{I}_0},
\end{align}
where $\mca{I}_0$ is quantum Fisher information for $\theta=0$ and $\ev{J}_\theta$ is the average current for the perturbed dynamics.

As shown in Appendix \ref{app:quantum.Fisher}, the quantum Fisher information $\mca{I}_0$ can be calculated as
\begin{equation}
	\mca{I}_0=4\eval{\partial_{\theta_1\theta_2}^2\qty(\ln|\tr\varrho_{\tau,\vb*{\theta}}|)}_{\theta_1=\theta_2=0},
\end{equation}
where $\varrho_{\tau,\vb*{\theta}}=\vec{\mca{T}}e^{\int_0^\tau \dd{t} \mca{L}_{t,\vb*{\theta}}}(\varrho_0)$ is the operator evolved according to the following modified superoperator:
\begin{align}
  \mca{L}_{t,\vb*{\theta}}(\circ)&=-i[H,\circ]+\sum_{k\geq 1}\sqrt{[1+\ell_{k}(t)\theta_1][1+\ell_{k}(t)\theta_2]}\mca{F}_k[ L_{k}\circ L_{k}^\dagger]\notag\\
  &-\frac{1}{2}\sum_{k\geq 1}[1+\ell_{k}(t)\theta_1]L_{k}^\dagger L_{k}\circ -\frac{1}{2}\sum_{k\geq 1}[1+\ell_{k}(t)\theta_2]\circ  L_{k}^\dagger L_{k},\label{eq:mod.op}
\end{align}
For simplicity of calculation, we introduce vectorization representation, which represents matrix $A=\sum_{m,n}a_{mn}\dyad{m}{n}$ as $\kvec{A}=\sum_{m,n}a_{mn}\ket{m}\otimes \ket{n}$.
By algebraic calculations, one can show that $\kvec{AB}=(A\otimes \mbb{1})\kvec{B}$ and $\kvec{BA}=(\mbb{1}\otimes A^\top )\kvec{B}$.
Note that a general CPTP map can be written as $\mca{F}_k[\circ]=\sum_{\alpha_k}K_k^{\alpha_k}\circ K_k^{\alpha_k \dagger}$, 
where $\sum_{\alpha_k}K_k^{\alpha_k \dagger}K_k^{\alpha_k} =\mbb{1}$.
Then, using vectorization representation, the superoperator $\mca{L}_{t,\vb*{\theta}}$ can be vectorized as
\begin{align}
  \widehat{\mca{L}}_{t,\vb*{\theta}}&\coloneqq -i(H\otimes \mbb{1}-\mbb{1}\otimes H^\top )
  +\sum_{k\geq 1}\sqrt{[1+\ell_{k}(t)\theta_1][1+\ell_{k}(t)\theta_2]}\sum_{\alpha_k}K_k^{\alpha_k} L_{k}\otimes (K_k^{\alpha_k}L_{k})^* \notag\\
  &-\frac{1}{2}\sum_{k\geq 1}[1+\ell_{k}(t)\theta_1]L_{k}^\dagger L_{k}\otimes \mbb{1} 
  -\frac{1}{2}\sum_{k\geq 1}[1+\ell_{k}(t)\theta_2]\mbb{1}\otimes (L_{k}^\dagger L_{k})^\top .
\end{align}
Therefore, we can calculate quantum Fisher information as follows:
\begin{align}
	\mca{I}_0&=4\eval{\qty[\partial_{\theta_1\theta_2}^2\bvec{\mbb{1}_S}\vec{\mca{T}} e^{\int_0^\tau \dd{t} \widehat{\mca{L}}_{t,\vb*{\theta}}}\kvec{\varrho_0} -\partial_{\theta_1}\bvec{\mbb{1}_S}\vec{\mca{T}}e^{\int_0^\tau \dd{t} \widehat{\mca{L}}_{t,\vb*{\theta}}}\kvec{\varrho_0} \partial_{\theta_2}\bvec{\mbb{1}_S}\vec{\mca{T}}e^{\int_0^\tau \dd{t} \widehat{\mca{L}}_{t,\vb*{\theta}}}\kvec{\varrho_0}]}_{\vb*{\theta}=\vb*{0}}\notag\\
	&=4\eval{\partial_{\theta_2}\bvec{\mbb{1}_S}\vec{\mca{T}}\int_0^\tau \dd{t} e^{\int_t^\tau \dd{s} \widehat{\mca{L}}_{s,\vb*{\theta}}}\partial_{\theta_1} \widehat{\mca{L}}_{t,\vb*{\theta}}e^{\int_0^t \dd{s} \widehat{\mca{L}}_{s,\vb*{\theta}}}\kvec{\varrho_0}}_{\vb*{\theta}=\vb*{0}}\notag\\
  &-4\eval{\bvec{\mbb{1}_S}\vec{\mca{T}}\int_0^\tau \dd{t} e^{\int_t^\tau \dd{s} \widehat{\mca{L}}_{s,\vb*{\theta}}}\partial_{\theta_1} \widehat{\mca{L}}_{t,\vb*{\theta}}e^{\int_0^t \dd{s} \widehat{\mca{L}}_{s,\vb*{\theta}}}\kvec{\varrho_0}\bvec{\mbb{1}_S}\vec{\mca{T}}\int_0^\tau \dd{t} e^{\int_t^\tau \dd{s} \widehat{\mca{L}}_{s,\vb*{\theta}}}\partial_{\theta_2} \widehat{\mca{L}}_{t,\vb*{\theta}}e^{\int_0^t \dd{s} \widehat{\mca{L}}_{s,\vb*{\theta}}}\kvec{\varrho_0}}_{\vb*{\theta}=\vb*{0}}\notag\\
  &=-4\eval{\bvec{\mbb{1}_S}\vec{\mca{T}}\int_0^\tau \dd{t} e^{\int_t^\tau \dd{s} \widehat{\mca{L}}_{s,\vb*{\theta}}}\partial_{\theta_1}\widehat{\mca{L}}_{t,\vb*{\theta}}e^{\int_0^t \dd{s} \widehat{\mca{L}}_{s,\vb*{\theta}}}\kvec{\varrho_0}\bvec{\mbb{1}_S}\vec{\mca{T}}\int_0^\tau \dd{t} e^{\int_t^\tau \dd{s} \widehat{\mca{L}}_{s,\vb*{\theta}}}\partial_{\theta_2} \widehat{\mca{L}}_{t,\vb*{\theta}}e^{\int_0^t \dd{s} \widehat{\mca{L}}_{s,\vb*{\theta}}}\kvec{\varrho_0}}_{\vb*{\theta}=\vb*{0}}\notag\\
  &+4\eval{\bvec{\mbb{1}_S}\vec{\mca{T}}\int_0^\tau\dd{t}\int_t^{\tau}\dd{t'} e^{\int_{t'}^\tau \dd{s} \widehat{\mca{L}}_{s,\vb*{\theta}}}\partial_{\theta_2} \widehat{\mca{L}}_{t',\vb*{\theta}}e^{\int_t^{t'} \dd{s} \widehat{\mca{L}}_{s,\vb*{\theta}}}\partial_{\theta_1} \widehat{\mca{L}}_{t,\vb*{\theta}}e^{\int_0^t \dd{s} \widehat{\mca{L}}_{s,\vb*{\theta}}}\kvec{\varrho_0}}_{\vb*{\theta}=\vb*{0}}\notag\\
  &+4\eval{\bvec{\mbb{1}_S}\vec{\mca{T}}\int_0^\tau\dd{t}\int_0^{t}\dd{t'} e^{\int_{t}^\tau \dd{s} \widehat{\mca{L}}_{s,\vb*{\theta}}}\partial_{\theta_1} \widehat{\mca{L}}_{t,\vb*{\theta}}e^{\int_{t'}^{t} \dd{s} \widehat{\mca{L}}_{s,\vb*{\theta}}}\partial_{\theta_2} \widehat{\mca{L}}_{t',\vb*{\theta}}e^{\int_0^{t'} \dd{s} \widehat{\mca{L}}_{s,\vb*{\theta}}}\kvec{\varrho_0}}_{\vb*{\theta}=\vb*{0}}\notag\\
  &+4\eval{\bvec{\mbb{1}_S}\vec{\mca{T}}\int_0^\tau \dd{t} e^{\int_t^\tau \dd{s} \widehat{\mca{L}}_{s,\vb*{\theta}}}\partial_{\theta_1\theta_2}^2 \widehat{\mca{L}}_{t,\vb*{\theta}}e^{\int_0^t \dd{s} \widehat{\mca{L}}_{s,\vb*{\theta}}}\kvec{\varrho_0}}_{\vb*{\theta}=\vb*{0}}\notag\\
  &=4\Big[-\bvec{\mbb{1}_S}\int_0^\tau \dd{t}\widehat{\mca{G}}_{1,t}\kvec{\varrho_t} \bvec{\mbb{1}_S}\int_0^\tau \dd{t}\widehat{\mca{G}}_{2,t}\kvec{\varrho_t}\notag\\
  &+\eval{\bvec{\mbb{1}_S}\vec{\mca{T}}\int_0^\tau\dd{t}\int_t^{\tau}\dd{t'} \widehat{\mca{G}}_{2,t'}e^{\int_t^{t'} \dd{s} \widehat{\mca{L}}_{s,\vb*{\theta}}}\widehat{\mca{G}}_{1,t}\kvec{\varrho_t}}_{\vb*{\theta}=\vb*{0}}\notag\\
  &+\eval{\bvec{\mbb{1}_S}\vec{\mca{T}}\int_0^\tau\dd{t}\int_0^{t}\dd{t'} \widehat{\mca{G}}_{1,t}e^{\int_{t'}^{t} \dd{s} \widehat{\mca{L}}_{s,\vb*{\theta}}}\widehat{\mca{G}}_{2,t'}\kvec{\varrho_{t'}}}_{\vb*{\theta}=\vb*{0}}\notag\\
  &+\eval{\bvec{\mbb{1}_S}\int_0^\tau \dd{t} \partial_{\theta_1\theta_2}^2 \widehat{\mca{L}}_{t,\vb*{\theta}}\kvec{\varrho_t}}_{\vb*{\theta}=\vb*{0}}
  \Big],
\end{align}
where we use the fact $\bvec{\mbb{1}_S}e^{\widehat{\mca{L}}t}=\bvec{\mbb{1}_S}$ and the operators $\widehat{\mca{G}}_{1,t}$ and $\widehat{\mca{G}}_{2,t}$ are defined as
\begin{align}
  \widehat{\mca{G}}_{1,t}&\coloneqq \eval{\partial_{\theta_1}\widehat{\mca{L}}_{t,\vb*{\theta}}}_{\vb*{\theta}
  =\vb*{0}}=\frac{1}{2}\sum_{k\geq 1}\ell_{k}(t)\qty[ \sum_{\alpha_k}K_k^{\alpha_k} L_{k}\otimes (K_k^{\alpha_k}L_{k})^* - (L_{k}^\dagger L_{k})\otimes\mbb{1}_S],\\
	\widehat{\mca{G}}_{2,t}&\coloneqq \eval{\partial_{\theta_2}\widehat{\mca{L}}_{t,\vb*{\theta}}}_{\vb*{\theta}=\vb*{0}}
  =\frac{1}{2}\sum_{k\geq 1}\ell_{k}(t)\qty[ \sum_{\alpha_k}K_k^{\alpha_k} L_{k}\otimes (K_k^{\alpha_k}L_{k})^* - \mbb{1}_S\otimes(L_{k}^\dagger L_{k} )^\top].
\end{align}
Note that $\bvec{\mbb{1}_S}\widehat{\mca{G}}_{1,t}\kvec{A}=\bvec{\mbb{1}_S}\widehat{\mca{G}}_{2,t}\kvec{A}=0$ holds for any operator $A$ because the map $\mca{F}_k$ is trace-preserving and
\begin{align}
  \bvec{\mbb{1}_S}\qty[ \sum_{\alpha_k}K_k^{\alpha_k} L_{k}\otimes (K_k^{\alpha_k}L_{k})^* - (L_{k}^\dagger L_{k})\otimes\mbb{1}_S]\kvec{A}
  &=\tr(\mca{F}_k[L_k AL_k^\dagger])-\tr(L_k AL_k^\dagger)=0.
\end{align}
Therefore, we obtain the following expression:
\begin{align}
  \mca{I}_0&=4\eval{\bvec{\mbb{1}_S}\int_0^\tau \dd{t} \partial_{\theta_1\theta_2}^2 \widehat{\mca{L}}_{t,\vb*{\theta}}\kvec{\varrho_t}}_{\vb*{\theta}=\vb*{0}}\notag\\
  &=\bvec{\mbb{1}_S}\int_0^\tau \dd{t} \sum_{k\geq 1}\ell_{k}^2(t)\sum_{\alpha_k}K_k^{\alpha_k} L_{k}\otimes (K_k^{\alpha_k}L_{k})^*\kvec{\varrho_t}\notag\\
  &=\int_0^\tau \dd{t} \sum_{k\geq 1}\ell_{k}^2 (t)\tr(L_{k}\varrho_t L_{k}^\dagger ).\label{eq:app.Fisher.eval}
\end{align}
Next, we evaluate $\mca{I}_0$ from above as 
\begin{align}
  \mca{I}_0
  &=\int_0^\tau \dd{t} \sum_{k\geq 1}\ell_{k}^2 (t)\tr(L_{k}\varrho_t L_{k}^\dagger )
  \notag\\
  &=\frac{1}{2}\int_0^\tau \dd{t} \sum_{k\geq 1}\qty[\ell_{k}^2 (t)\tr(L_{k}\varrho_t L_{k}^\dagger )
  +\ell_{k^*}^2 (t)\tr(L_{k^*}\varrho_t L_{k^*}^\dagger )]
  \notag\\
  &=\frac{1}{2}\int_0^\tau \dd{t}\sum_{k\geq 1}\ell_{k}^2 (t)
  \qty[\tr(L_{k}\varrho_t L_{k}^\dagger )+\tr(L_{k^*}\varrho_t L_{k^*}^\dagger )]
  \notag\\
  &=\frac{1}{2}\int_0^\tau \dd{t} \sum_{k\geq 1}
  \dfrac{ \qty[\tr(L_{k} \varrho_t L_{k}^\dagger) - \tr(L_{k^*} \varrho_t L_{k^*}^\dagger) ]^2}
  {\tr(L_{k} \varrho_t L_{k}^\dagger) + \tr(L_{k^*} \varrho_t L_{k^*}^\dagger) }
  \notag\\
  &=\frac{1}{2}\int_0^\tau \dd{t} \sum_{k\geq 1}\dfrac{ \qty[\sum_{m,n}(w_{k}^{mn}p_n-w_{k^*}^{nm}p_m)]^2}{\sum_{m,n}(w_{k}^{mn}p_n+w_{k^*}^{nm}p_m)}
  \notag\\
  &\le\frac{1}{2}\int_0^\tau \dd{t} \sum_{k\geq 1,m,n}\dfrac{ \qty[(w_{k}^{mn}p_n-w_{k^*}^{nm}p_m)]^2}{w_{k}^{mn}p_n+w_{k^*}^{nm}p_m}
  \notag\\
  &= \int_0^\tau \dd{t} \sum_{k\geq 1}\sum_{m,n}
  \frac{(\sigma^{mn}_k)^2}{8a^{mn}_k} \Phi \qty(\frac{\sigma^{mn}_k}{2a^{mn}_k})^{-2}
  \notag\\
  &\leq \int_0^\tau \dd{t} 
  \frac{{\sigma}^2}{4\dot{A}} \Phi \qty(\frac{{\sigma}}{2\dot{A}})^{-2},
\end{align}
where we apply the Cauchy-Schwarz inequality to obtain the sixth line and Jensen's inequality for the concave function $(x^2/y)\Phi(x/y)^{-2}$ to obtain the last line.
In addition, since $x^2\Phi(x)^{-2}$ is a monotonically increasing function (see Proposition \ref{prop:mono.inc}) and ${\sigma}\leq \dot{\Sigma}$ holds, we have
\begin{equation}
    \frac{{\sigma}^2}{4\dot{A}} \Phi \qty(\frac{{\sigma}}{2\dot{A}})^{-2}\leq
\frac{\dot{\Sigma}^2}{4\dot{A}} \Phi \qty(\frac{\dot{\Sigma}}{2\dot{A}})^{-2}.
\end{equation}
Therefore, applying Jensen's inequality again yields 
\begin{align}
  \mca{I}_0 &\leq 
  \int_0^\tau \dd{t} \frac{\dot{\Sigma}^2}{4\dot{A}} \Phi \qty(\frac{\dot{\Sigma}}{2\dot{A}})^{-2}
  \\
  &\leq 
  \frac{\Sigma^2}{4A} \Phi \qty(\frac{\Sigma}{2A})^{-2}.\label{eq:Fi.tkur}
\end{align}

Finally, we evaluate ${\partial_{\theta}\ev{J}_\theta}$ at $\theta=0$.
By expressing the density operator in powers of $\theta$ as ${\varrho}_{t,\theta}={\varrho}_t + \theta {\varphi}_t+O(\theta^2)$, the perturbed dynamics can be rewritten as
\begin{align}
  \dot{\varrho}_t + \theta \dot{\varphi}_t &= -i \left[H, \varrho_t + \theta \varphi_t \right] 
  + \sum_{k\geq 1} [1 + \ell_{k}(t) \theta] \qty(\mca{F}_k[L_k(\varrho_t + \theta \varphi_t)L_k^\dagger]-\frac{1}{2}\{L_k^\dagger L_k,\varrho_t + \theta \varphi_t \})
  +O(\theta^2).
\end{align}
Then, we obtain the following equation by collecting the first-order terms:
\begin{align}
	\dot{\varphi}_t &=-i[H,\varphi_t]+\sum_{k\geq 1}\qty(\mca{F}_k[L_k\varphi_t L_k^\dagger]-\frac{1}{2}\{L_k^\dagger L_k,\varphi_t \})
  +\sum_{k\geq 1}\ell_{k}(t)\qty(\mca{F}_k[L_k\varrho_t L_k^\dagger]-\frac{1}{2}\{L_k^\dagger L_k,\varrho_t \})
  \notag\\
  &=\mca{L}^{\mrm{(fb)}}[\varphi_t]+\sum_{k\geq 1}\ell_{k}(t)\qty(\mca{F}_k[L_k\varrho_t L_k^\dagger]-\frac{1}{2}\{L_k^\dagger L_k,\varrho_t \}),
\end{align}
where the initial condition is $\varphi_0=\mbb{0}$.
Noting that $c_{k}=-c_{k^*}$ and $c_{k}\ell_k=c_{k^*}\ell_{k^*}$, the partial derivative of the current average in the perturbed dynamics with respect to $\theta$ can be calculated as
\begin{align}
	\eval{\partial_{\theta}\ev{J}_\theta}_{\theta=0}
  &=\eval{\partial_{\theta}\qty[\int_0^\tau\dd{t}\sum_{k\geq 1}c_{k}[1+\ell_{k}(t)\theta]\tr{L_{k}(\varrho_t+\theta\varphi_t)L_{k}^\dagger}+O(\theta^2)]}_{\theta=0}\notag\\
	&=\int_0^\tau\dd{t}\sum_{k\geq 1}c_{k}\ell_{k}(t)\tr(L_{k}\varrho_t L_{k}^\dagger)+\int_0^\tau\dd{t}\sum_{k\geq 1}c_{k}\tr(L_{k}\varphi_t L_{k}^\dagger)\notag\\
	&=\frac{1}{2}\int_0^\tau\dd{t}\sum_{k\geq 1}c_{k}\qty[\tr(L_{k}\varrho_t L_{k}^\dagger)-\tr(L_{k^*}\varrho_t L_{k^*}^\dagger)]+\int_0^\tau\dd{t}\sum_{k\geq 1}c_{k}\tr(L_{k}\varphi_t L_{k}^\dagger)\notag\\
	&=\int_0^\tau\dd{t}\sum_{k\geq 1}c_{k}\tr(L_{k}\varrho_t L_{k}^\dagger)+\int_0^\tau\dd{t}\sum_{k\geq 1}c_{k}\tr(L_{k}\varphi_t L_{k}^\dagger)\notag\\
	&=\ev{J}+\ev{J}_\varphi\notag\\
    &=\ev{J}(1+\delta_J),\label{eq:app.par.avg.TUR}
\end{align}
where $\ev{J}_\varphi\coloneqq \int_0^\tau\dd{t}\sum_{k\geq 1}c_{k}\tr(L_{k}\varphi_t L_{k}^\dagger)$ and $\delta_J={\ev{J}_\varphi}/{\ev{J}}$.

Finally, combining relations \eqref{eq:qCRcound}, \eqref{eq:Fi.tkur}, and \eqref{eq:app.par.avg.TUR} leads to the improved TUR [Eq.~(\TKUR) in the main text]:
\begin{align}
  \frac{\Var[J]}{\ev{J}^2}
  \ge (1+\delta_{J})^2\frac{4A}{\Sigma^2}\Phi\qty(\frac{\Sigma}{2A})^2.
\end{align}
By further applying the inequality $\Phi(x)\ge \sqrt{x}$, the main result (\FirRes) is immediately obtained:
\begin{align}
  \frac{\Var[J]}{\ev{J}^2}\ge (1+\delta_{J})^2\frac{4A}{\Sigma^2}\Phi\qty(\frac{\Sigma}{2A})^2
  \ge \frac{2(1+\delta_{J})^2}{\Sigma}.
\end{align}

\section{Special cases}
In this section, we examine several special cases of our general framework.

\subsection{Absence of feedback}
We first consider the case without feedback, i.e., $\mathcal{F}_k[\circ] = \circ$ for all $k$. In this situation, each $\mathcal{F}_k$ trivially satisfies the condition of a unital CPTP map, and thus the setting lies within the scope of our results. In particular, this case is included in unitary feedback, and hence $\dot{\Sigma} = \sigma$ holds.
In the following, we show that our result reduces to the previously established thermokinetic uncertainty relation (TKUR) \cite{Vu.2025.PRXQ} in this limit.

The master equation of Markovian feedback dynamics becomes
\begin{align}
  \dot{\varrho}_t
  = -i[H,\varrho_t]
    + \sum_{k \ge 1} \mathcal{D}[L_k]\varrho_t,
\end{align}
which is exactly the GKSL equation.
The evolution equation for $\varphi_t$ reads
\begin{align}
  \dot{\varphi}_t
  = -i[H,\varphi_t]
    + \sum_{k \ge 1} \mathcal{D}[L_k]\varphi_t
    + \sum_{k \ge 1} \ell_k \mathcal{D}[L_k]\varrho_t.
\end{align}
Thus, both $\varrho_t$ and $\varphi_t$ obey the same evolution as in Ref.~\cite{Vu.2025.PRXQ}.
Next, consider the change in the mutual information,
$dI = I_{t+dt} - I_{t+dt}^-$. In the absence of feedback, we have
$\varrho_{t+dt} = \varrho_{t+dt}^-$ and
$\varrho_{t+dt,k} = \varrho_{t+dt,k}^-$.
Therefore, $I_{t+dt} = I_{t+dt}^-$, and hence $dI = 0$.
Consequently, in this setting, our TKUR (\TKUR) coincides exactly with that derived in Ref.~\cite{Vu.2025.PRXQ} in the absence of feedback.

\subsection{Short-time limit}
We consider the short-time limit and show that $\delta_J = O(\tau)$; hence, $\delta_J$ vanishes as $\tau \to 0$.
The evolution equation for $\varphi_t$ is given by
\begin{align}
    \dot{\varphi}_t &=-i[H,\varphi_t]+\sum_{k\geq 1}\qty(\mca{F}_k[L_k\varphi_t L_k^\dagger]-\frac{1}{2}\{L_k^\dagger L_k,\varphi_t \})
  +\sum_{k\geq 1}\ell_{k}(t)\qty(\mca{F}_k[L_k\varrho_t L_k^\dagger]-\frac{1}{2}\{L_k^\dagger L_k,\varrho_t \})
  \notag\\
  &=\mca{L}^{\mrm{(fb)}}[\varphi_t]+\sum_{k\geq 1}\ell_{k}(t)\qty(\mca{F}_k[L_k\varrho_t L_k^\dagger]-\frac{1}{2}\{L_k^\dagger L_k,\varrho_t \}),
\end{align}
We expand $\varphi_\tau$ to first order in $\tau$. Using a short-time expansion, we obtain
\begin{align}
  \varphi_{\tau}&=\varphi_0
  +\tau \qty{\mca{L}^{\mrm{(fb)}}[\varphi_0] + \sum_{k\geq 1}\ell_{k}(0)\qty(\mca{F}_k[L_k\varrho_0 L_k^\dagger]-\frac{1}{2}\{L_k^\dagger L_k,\varrho_0 \})}
  +O(\tau^2)
  \notag\\
  &=\tau \sum_{k\geq 1}\ell_{k}(0)\qty(\mca{F}_k[L_k\varrho_0 L_k^\dagger]-\frac{1}{2}\{L_k^\dagger L_k,\varrho_0 \})
  +O(\tau^2)
  \notag\\
  &=O(\tau),
\end{align}
where we use the initial condition $\varphi_0 = \mbb{0}$.
Consequently, $\ev{J}_\varphi$ can be evaluated as
\begin{align}
  \ev{J}_\varphi
  &= \int_0^\tau dt \sum_{k\geq 1} c_k \tr(L_k\varphi_t L_k^\dagger)
  \notag\\
  &= O(\tau^2).
\end{align}
On the other hand, the current average scales as $\ev{J} \propto \tau$ in the short-time limit, provided $\sum_{k\ge 1}c_k\tr(L_k\varrho_0 L_k^\dagger)\neq 0$. Therefore,
\begin{equation}
  \delta_J = \frac{\ev{J}_\varphi}{\ev{J}} = O(\tau),
\end{equation}
which vanishes as $\tau \to 0$.

\section{Numerical demonstrations}\label{sup:num}
\subsection{Two-level quantum clock}
\begin{figure}[t]
	\centering
	\includegraphics[width=1\linewidth]{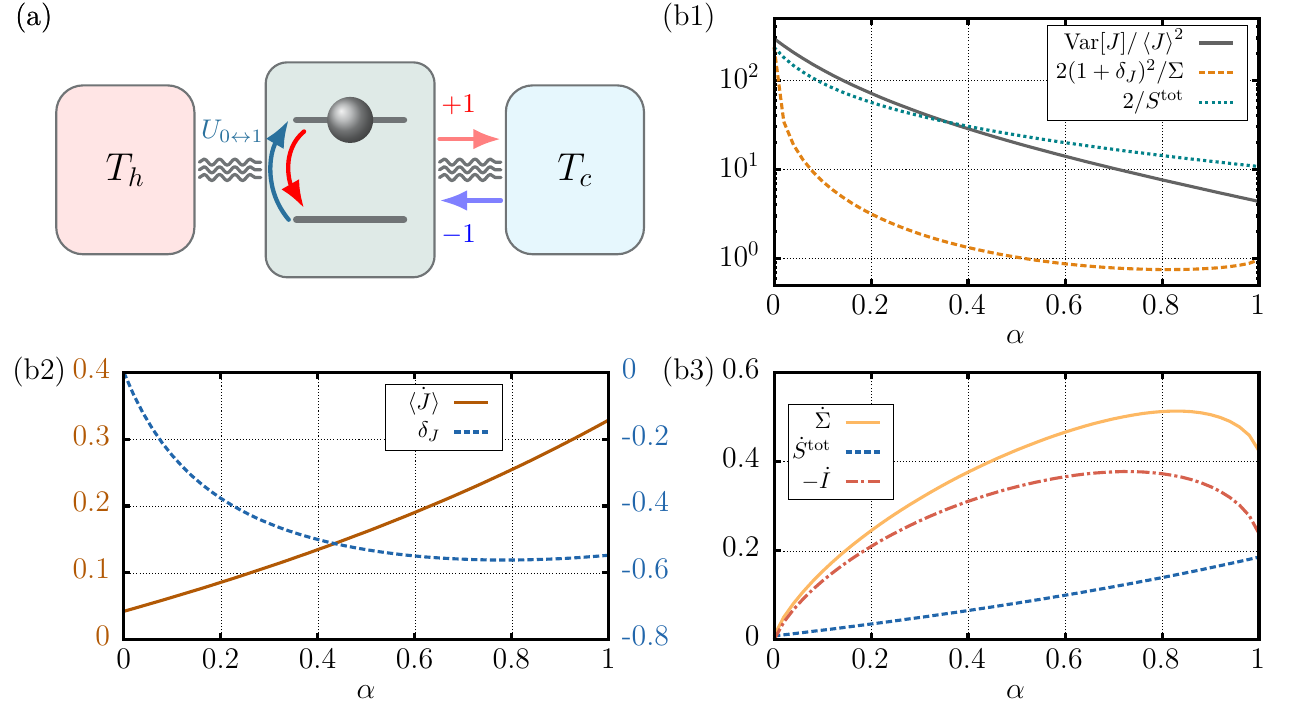}
	\protect\caption{(a) Schematic illustration of the two-level quantum clock weakly coupled to two thermal reservoirs. (b) Numerical verification of the quantum TUR (\FirRes) under feedback in the stationary state. The vertical axis in (b1) is plotted on a logarithmic scale, while all other axes are linear. In (b1), the solid, dashed, and dotted lines represent the relative fluctuation $\Var[J]/\ev{J}^2$, its lower bound $2(1+\delta_{J})^2/\Sigma$, and the original TUR lower bound $2/S^{\rm{tot}}$, respectively. In (b2), the solid and dashed lines show the average current rate $\ev{\dot J}$ and the correction term $\delta_{J}$, respectively. In (b3), the solid, dashed, and dash-dotted lines represent $\dot\Sigma$, $\dot S^{\rm tot}$, and $-\dot I$, respectively. The horizontal axes indicate the feedback weight $\alpha$ for the jump $L_{1\rightarrow 0}^{(h)}$. The remaining parameters are fixed at $E_0=0$, $E_1=1$, $\beta_h =0.2$, $\beta_c=0.4$, $\gamma_{1\rightarrow 0}^{(h)}=\gamma_{1\rightarrow 0}^{(c)}=1$, and $\tau = 1$.}
\label{fig:two.reservoir.clock}
\end{figure}

To illustrate the smooth connection between our result [Eq.~(\FirRes)] and the original TUR, we consider a two-level clock model with Hamiltonian $H=\sum_{i=0}^{1}E_i \dyad{i}$, where $E_i$ denotes the energy of state $\ket{i}$.
The clock is simultaneously coupled to two thermal reservoirs: a hot reservoir at temperature $T_h=1/\beta_h$ and a cold reservoir at temperature $T_c=1/\beta_c$ [see Fig.~\ref{fig:two.reservoir.clock}(a)].
Both reservoirs induce transitions between the states $\ket{0}$ and $\ket{1}$.
The jump operator associated with the transition from $\ket{i}$ to $\ket{j}$ mediated by reservoir $\nu$ is $L_{i\rightarrow j}^{(\nu)}=\sqrt{\gamma_{i\rightarrow j}^{(\nu)}}\dyad{j}{i}$, where $\gamma_{i\rightarrow j}^{(\nu)}$ is the corresponding transition rate.
These rates satisfy the local detailed-balance condition $\gamma_{i\rightarrow j}^{(\nu)}=e^{-\beta_\nu(E_j-E_i)}\gamma_{j\rightarrow i}^{(\nu)}$.
In the absence of feedback, the dynamics is purely classical and obeys a classical master equation; hence, the original TUR applies.

We introduce a unital feedback operation $\mca{F}[\circ]=\alpha U_{0\leftrightarrow 1}\circ U_{0\leftrightarrow 1}^{\dagger} +(1-\alpha)\circ$ after the jump $L_{1\rightarrow 0}^{(h)}$, while no feedback is applied after the other jumps.
Here, $U_{0\leftrightarrow 1}=\dyad{0}{1}+\dyad{1}{0}$ is the unitary operator that swaps the states $\ket{0}$ and $\ket{1}$, and $\alpha\in[0,1]$ denotes the feedback weight.
For $\alpha=1$, the swap operation is applied after every occurrence of the jump $L_{1\rightarrow 0}^{(h)}$, whereas $\alpha=0$ corresponds to the no-feedback limit.
Consequently, when $\alpha=1$, hot-reservoir-induced transitions effectively bring the system into the state $\ket{1}$.
Under this feedback protocol, the clock dynamics is governed by
\begin{align}
  \dot{\varrho}_{t}&=-i[H, \varrho_t]
  +\mca{F}[L_{1\rightarrow 0}^{(h)}\varrho_t (L_{1\rightarrow 0}^{(h)})^\dagger]-\{(L_{1\rightarrow 0}^{(h)})^\dagger L_{1\rightarrow 0}^{(h)},\varrho_t\}/2
  +\mca{D}[L_{0\rightarrow 1}^{(h)}]\varrho_t
  +\mca{D}[L_{1\rightarrow 0}^{(c)}]\varrho_t
  +\mca{D}[L_{0\rightarrow 1}^{(c)}]\varrho_t \notag\\
  &=-i[H, \varrho_t]
  +\alpha \mca{D}[U_{0\leftrightarrow 1}L_{1\rightarrow 0}^{(h)}]\varrho_t
  +(1-\alpha)\mca{D}[L_{1\rightarrow 0}^{(h)}]\varrho_t
  +\mca{D}[L_{0\rightarrow 1}^{(h)}]\varrho_t
  +\mca{D}[L_{1\rightarrow 0}^{(c)}]\varrho_t
  +\mca{D}[L_{0\rightarrow 1}^{(c)}]\varrho_t.
\end{align}
Here, $\mca{D}[L]\circ\coloneqq L\circ L^\dagger-\{L^\dagger L,\circ\}/2$.
We emphasize that, owing to the presence of two distinct reservoirs, a persistent current exists even in the absence of feedback ($\alpha=0$), in contrast to the case of a single-reservoir quantum clock.

In the following, we focus on the tick current $J$, which assigns $+1$ to each occurrence of the jump $L_{1\rightarrow 0}^{(c)}$, $-1$ to each occurrence of $L_{0\rightarrow 1}^{(c)}$, and $0$ to all other transitions.
Thus, this current measures the net number of ticks generated by the clock.
To demonstrate that our result, Eq.~(\FirRes), continuously recovers the original TUR in the classical no-feedback limit, we numerically evaluate all relevant quantities in the steady state.
Motivated by experimentally relevant situations, we analyze the clock dynamics over a finite observation time $\tau$, rather than in the asymptotic long-time limit.
We vary the feedback weight $\alpha$ from $0$ to $1$, while keeping all other parameters fixed.
The results are presented in Fig.~\ref{fig:two.reservoir.clock}(b).

Figures~\ref{fig:two.reservoir.clock}(b1) and \ref{fig:two.reservoir.clock}(b2) show that the relative fluctuation $\Var[J]/\ev{J}^{2}$ decreases, while the average current $\ev{J}$ increases, as the feedback weight $\alpha$ is increased. This indicates that feedback improves the precision of the clock current.
Figure~\ref{fig:two.reservoir.clock}(b1) further shows that the relative fluctuation is always bounded from below by $2(1+\delta_J)^2/\Sigma$, providing numerical support for our main result, Eq.~(\FirRes).
For $\alpha=0$, Figs.~\ref{fig:two.reservoir.clock}(b2) and \ref{fig:two.reservoir.clock}(b3) confirm that $\delta_J=0$ and $\dot I=0$, respectively.
In this limit, our TUR reduces to the previously established quantum TUR \cite{Vu.2025.PRXQ}.
Since the present model becomes classical in the absence of feedback, it further reduces to the original TUR, $\Var[J]/\ev{J}^{2} \ge 2/S^{\rm tot}$ \cite{Gingrich.2016.PRL}.
We also find that the quantity $2/S^{\rm tot}$ alone, obtained by neglecting both the mutual-information contribution and the correction term $\delta_J$, ceases to be a valid lower bound when $\alpha>0.4$.
This observation demonstrates the necessity of the additional feedback-induced terms involving $I$ and $\delta_J$ in our result.
Finally, Fig.~\ref{fig:two.reservoir.clock}(b3) shows that $\dot{\Sigma}=\dot{S}^{\rm tot}-\dot{I}$ remains nonnegative throughout the parameter range considered, thereby confirming the second-law-like inequality, Eq.~(\SecRes), for this model.

\subsection{Three-level quantum clock}
To further validate our results, we numerically investigate the three-level quantum clock model introduced in the main text in a different parameter regime.
The model is identical to that considered in the main text, except that we vary the jump rate $\gamma_{1\rightarrow 0}$ while keeping all other parameters fixed.

Since the feedback operation is triggered by jumps between $\ket{0}$ and $\ket{1}$, the effective feedback strength is controlled by $\gamma_{1\rightarrow 0}$.
For small $\gamma_{1\rightarrow 0}$, feedback events occur rarely, and their influence is weak, whereas larger values of $\gamma_{1\rightarrow 0}$ lead to stronger feedback effects.

\begin{figure}[t]
	\centering
	\includegraphics[width=1\linewidth]{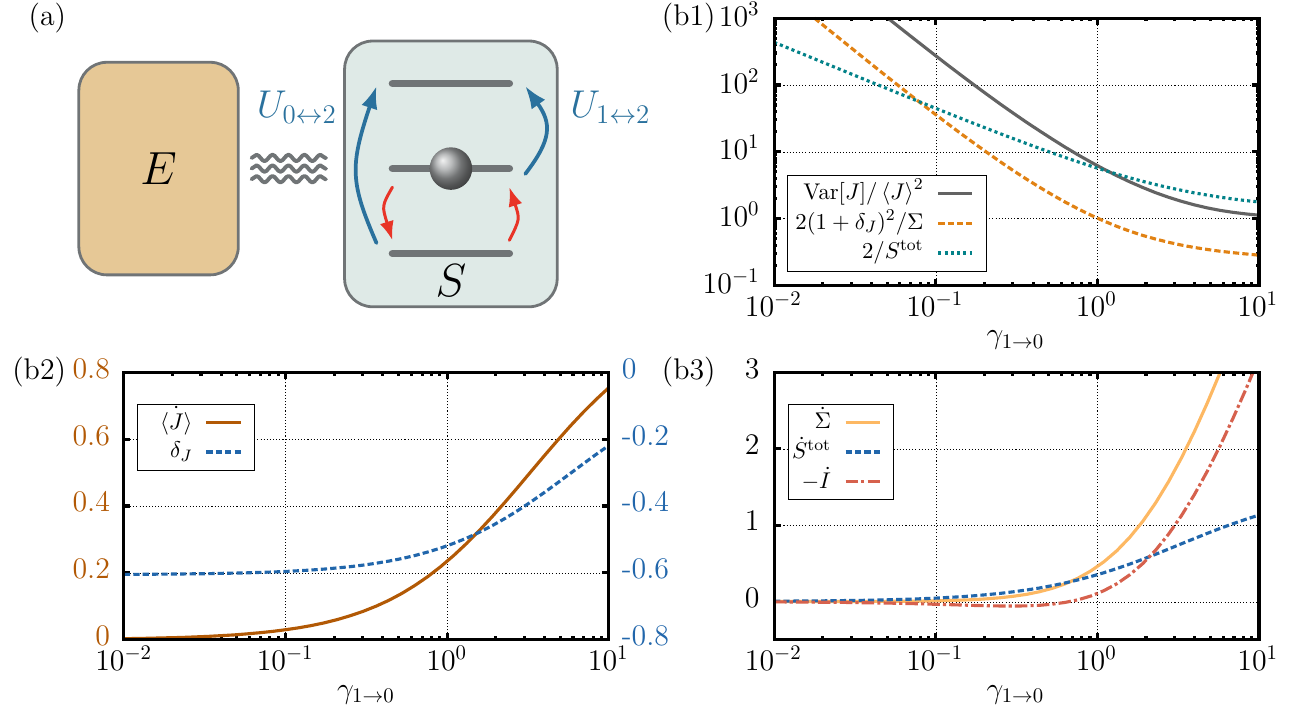}
	\protect\caption{(a) Schematic illustration of the three-level quantum clock, weakly coupled to a thermal reservoir. (b) Numerical verification of the quantum TUR (\FirRes) with feedback in the stationary state. The vertical axis of (b1) and the horizontal axes of (b1)--(b3) are shown on a logarithmic scale, while all other axes are linear. (b1) Solid, dashed, and dotted lines represent the relative fluctuation $\Var[J]/\ev{J}^2$, its lower bound $2(1+\delta_{J})^2/\Sigma$, and $2/S^{\mrm{tot}}$, respectively. (b2) Solid and dashed lines depict the average current rate $\dot{\ev{J}}$ and the correction term $\delta_{J}$, respectively. (b3) Solid, dashed, and dash-dotted lines represent $\dot\Sigma$, $\dot S^{\rm tot}$, and $-\dot I$, respectively. The jump rate of $L_{1\rightarrow 0}$, $\gamma_{1\rightarrow 0}$, is varied over the range $[10^{-2},10]$, while the remaining parameters are fixed at $E_0=0$, $E_1=0.5$, $E_2=1$, $\beta =1$, $\gamma_{2\rightarrow 1}=1$, $\gamma_{2\rightarrow 0}=1$, and $\tau = 1$.}
\label{fig:clock2}
\end{figure}

The numerical results are shown in Fig.~\ref{fig:clock2}.
For $\gamma_{1\rightarrow 0}=0$, the average current, total entropy production, and information contribution $I$ all vanish, as shown in Figs.~\ref{fig:clock2}(b2) and \ref{fig:clock2}(b3).
At the same time, Fig.~\ref{fig:clock2}(b1) shows that the relative fluctuation of the current diverges.
These behaviors arise because, when $\gamma_{1\rightarrow 0}=0$, no feedback events occur, and the system relaxes to an equilibrium steady state.
The vanishing average current then causes the relative fluctuation $\Var[J]/\ev{J}^2$ to diverge.

Figure~\ref{fig:clock2}(b1) also shows that the quantity $2/S^{\rm tot}$ fails to provide a lower bound on the relative fluctuation in the presence of feedback.
This demonstrates the necessity of including both the information contribution $I$ and the correction term $\delta_J$ in our TUR.
Moreover, Fig.~\ref{fig:clock2}(b3) shows that the information contribution rate $-\dot I$ increases for $\gamma_{1\rightarrow 0}>1$, reflecting the enhanced influence of feedback at larger jump rates.

We further note that $\delta_J=\ev{J}_\varphi/\ev{J}$ remains finite in the limit $\gamma_{1\rightarrow 0}\to0$ [Fig.~\ref{fig:clock2}(b2)], because $\ev{J}$ and $\ev{J}_\varphi$ vanish at the same order in $\gamma_{1\rightarrow 0}$.
This can be understood as follows.
For arbitrarily small but nonzero $\gamma_{1\rightarrow0}$, the feedback effect is infinitesimal yet finite, perturbing the steady state slightly away from equilibrium and giving rise to a small but finite current $\ev{J}$.
The feedback-induced quantity $\ev{J}_\varphi$ originates from the same infinitesimal effect and is therefore of the same order, so their ratio $\delta_J$ stays finite even as both currents vanish.
At the singular point $\gamma_{1\rightarrow 0}=0$, however, feedback is absent, and both vanish exactly, $\ev{J}=\ev{J}_\varphi=0$.

\section{Propositions}\label{sup:prop}

\begin{proposition}\label{prop:full.rank}
For any invertible (full-rank) density operators $\varrho$ and $\varrho'=\varrho+\lambda\varrho^{(1)}+O(\lambda^2)$, the von Neumann entropy of $\varrho'$ can be calculated as
\begin{equation}
    S(\varrho')=S(\varrho)-\lambda\tr(\varrho^{(1)}\ln\varrho) - \lambda\tr\varrho^{(1)} + O(\lambda^2).
\end{equation}
\end{proposition}
\begin{proof}
First, for any positive-definite operator $X$, we have the following identity:
\begin{equation}
    \ln X=\int_0^\infty\dd{s}\qty[\frac{\mbb{1}}{1+s}-(X+s\mbb{1})^{-1}].
\end{equation}
This can be easily proved by using the spectral decomposition of $X$ and performing basic integrations.
Next, using the identity $X^{-1}-Y^{-1}=X^{-1}(Y-X)Y^{-1}$ for any invertible operators $X$ and $Y$, it follows that
\begin{equation}
    [\varrho+\lambda\varrho^{(1)}+O(\lambda^2) + s\mbb{1}]^{-1} = (\varrho + s\mbb{1})^{-1} - \lambda(\varrho + s\mbb{1})^{-1}\varrho^{(1)}(\varrho + s\mbb{1})^{-1} + O(\lambda^2).
\end{equation}
Using these expressions, we can calculate the von Neumann entropy of $\varrho'$ as follows:
\begin{align}
    S(\varrho')&=-\tr(\varrho'\ln\varrho')\notag\\
    &=-\tr\int_0^\infty\dd{s}[\varrho+\lambda\varrho^{(1)}]\qty[\frac{\mbb{1}}{1+s}-(\varrho + s\mbb{1})^{-1} + \lambda(\varrho + s\mbb{1})^{-1}\varrho^{(1)}(\varrho + s\mbb{1})^{-1}] + O(\lambda^2)\notag\\
    &=S(\varrho)-\lambda\tr(\varrho^{(1)}\ln\varrho) - \lambda\tr\varrho^{(1)} + O(\lambda^2).
\end{align}
Here, we use the fact that
\begin{equation}
    \int_0^\infty\dd{s}(\varrho+s\mbb{1})^{-1}\varrho(\varrho+s\mbb{1})^{-1}=\mbb{1}.
\end{equation}
\end{proof}

\begin{proposition}\label{prop:unital}
The unitality of the CPTP map $\mathcal{F}[\circ]$ (i.e., $\mathcal{F}[\mbb{1}]=\mbb{1}$) is equivalent to the inequality $S(\mathcal{F}[\varrho]) - S(\varrho) \ge 0$ holding for any density operator $\varrho$.
\end{proposition}
\begin{proof}
First, we show that $S(\mathcal{F}[\varrho]) - S(\varrho)\ge 0$ for any density operator $\varrho$ if $\mathcal{F}[\mbb{1}]=\mbb{1}$.
Since the relative entropy $D(\varrho\|\phi)=\tr[\varrho(\ln\varrho-\ln\phi)]$ is monotonically decreasing under CPTP maps, we have
\begin{align}
  D(\mathcal{F}[\varrho] \| \mathcal{F}[\mbb{1}/d])
  \le
  D(\varrho \| \mbb{1}/d) .
\end{align}
Using $\mathcal{F}[\mbb{1}]=\mbb{1}$ and $\tr\varrho=\tr\mathcal{F}[\varrho]=1$, this inequality becomes
\begin{align}
  -S(\mathcal{F}[\varrho]) + \ln d
  \le
  -S(\varrho) + \ln d ,
\end{align}
which immediately yields $S(\mathcal{F}[\varrho]) - S(\varrho) \ge 0$.
Here, $d$ denotes the dimension of the Hilbert space.

Next, we show that $\mathcal{F}[\mbb{1}]=\mbb{1}$ holds if $S(\mathcal{F}[\varrho]) \ge S(\varrho)$ for any density operator $\varrho$.
Note that the von Neumann entropy attains its maximum only at the maximally mixed state $\varrho=\mbb{1}/d$, for which $S(\varrho)=\ln d$.
Therefore, $S(\varrho)\le \ln d$ for any $\varrho$.
Since $S(\mathcal{F}[\mbb{1}/d]) \ge S(\mbb{1}/d) = \ln d$, it is obvious that $S(\mathcal{F}[\mbb{1}/d])=\ln d$.
Consequently, $\mathcal{F}[\mbb{1}/d] = \mbb{1}/d$, or equivalently, $\mathcal{F}[\mbb{1}] = \mbb{1}$.
\end{proof}

\begin{proposition}\label{prop:mono.inc}
$x^2\Phi(x)^{-2}$ is a monotonically increasing function for $x>0$.
\end{proposition}
\begin{proof}
It suffices to show that $x\Phi(x)^{-1}$ is a monotonically increasing function.
Since $x\tanh(x)$ is strictly increasing on $(0,\infty)$, its inverse function $\Phi$ is also strictly increasing.
Let $z=\Phi(x)$, so that $x=z\tanh(z)$.
Then, $x\Phi(x)^{-1}=x/z=\tanh(z)=\tanh[\Phi(x)]$.
Because both $\tanh$ and $\Phi$ are increasing functions, it follows that $x\Phi(x)^{-1}$ is monotonically increasing.
\end{proof}

%\bibliography{refs}
